\documentclass[onecolumn,usegraphicx]{mn2e}

\usepackage[fleqn]{amsmath}
\usepackage{color}
\usepackage{times} 
\usepackage{amssymb}
\usepackage[outline]{contour}
\pagerange{001--0\pageref{lastpage}}

\newcommand{\beq}{\begin{equation}}
\newcommand{\eeq}{\end{equation}}
\newcommand{\bea}{\begin{eqnarray}}
\newcommand{\eea}{\end{eqnarray}}

\newcommand{\ii}{{\rm{i}}}
\newcommand{\fett}[1]{\boldsymbol{#1}}
\newcommand{\nabx}{\boldsymbol{\nabla}_{\boldsymbol{x}}}

\newcommand{\be}{\begin{equation}}
\newcommand{\ee}{\end{equation}}

\newcommand{\nab}{\fett{\nabla}}
\newcommand{\nabr}{\fett{\nabla}_{\fett{r}}}
\newcommand{\nabL}[1]{\nabla_{#1}^{\rm L}}
\newcommand{\nabLT}{\nab^{\rm L}}

\definecolor{darkgreen}{rgb}{0,0.5,0}

\newcommand{\rf}[1]{(\ref{#1})}

\contourlength{0.4pt} 
\contournumber{10} 

\makeatletter
\newcommand{\superbig}{\bBigg@{4}}
\newcommand{\megabig}{\bBigg@{5}}
\makeatother

\begin{document}

\title[How smooth are particle trajectories in a $\Lambda$CDM Universe?]{How smooth are particle trajectories in a \contour{black}{$\Lambda$}CDM Universe?}

\author[Cornelius Rampf, Barbara Villone and Uriel Frisch]
       {Cornelius Rampf,$^{1,2,3}$\thanks{rampf@physik.rwth-aachen.de} Barbara
Villone$^{2,4}$ and Uriel Frisch$^2$ \\
$^1$ Institute of Cosmology and Gravitation, University of Portsmouth,
Portsmouth PO1 3FX, United Kingdom \\
$^2$ Lab. Lagrange, UCA, OCA, CNRS, CS 34229, F-06304 Nice Cedex 4, France \\
$^3$ Max-Planck-Institute for Gravitational Physics (Albert-Einstein-Institute), D-14476 Potsdam-Golm, Germany \\
$^4$ INAF, Osservatorio Astrofisico di Torino, Via Osservatorio, 20, I-10025 Pino
Torinese, Torino, Italy 
}

\maketitle

\begin{abstract}
It is shown here that in a flat, cold dark matter (CDM) dominated Universe
with positive cosmological constant ($\Lambda$), modelled in terms of
a Newtonian and collisionless fluid, particle trajectories are
analytical in time (representable by a convergent Taylor series) until
at least a finite time after decoupling. 
The time variable used for this statement is the cosmic scale
factor, i.e., the ``$a$-time'', and not the cosmic time.
For this, a Lagrangian-coordinates formulation of the Euler--Poisson equations is employed, 
originally used by Cauchy for \mbox{3-D} incompressible flow.
Temporal analyticity for $\Lambda$CDM is found to be a consequence of
novel explicit all-order recursion relations
for the $a$-time Taylor coefficients of the Lagrangian displacement field,
from which we derive the convergence of the $a$-time Taylor series.
A lower bound for
the $a$-time where analyticity is guaranteed and shell-crossing is
ruled out is obtained, whose value depends only on $\Lambda$ and on
the initial spatial smoothness of the density field. The largest time interval
is achieved when $\Lambda$ vanishes, i.e., for an Einstein--de Sitter
universe.
Analyticity holds also if, instead of the $a$-time, one uses the
linear structure growth $D$-time, but no simple recursion relations
are then obtained. The analyticity result also holds when a curvature term 
is included in the Friedmann equation for the background, but inclusion 
of a radiation term arising from the primordial era spoils analyticity. 
\end{abstract}

\begin{keywords}
cosmology:~theory -- large scale structure of Universe -- dark matter
-- dark energy
\end{keywords}


\section{Introduction}

Recent observational results (Planck Collaboration XVI 2014) indicate that we live in a spatially
flat $\Lambda$CDM Universe
which is nowadays dominated by a cold dark matter (CDM) component
and a cosmological constant, $\Lambda >0$.
In a primordial era, matter was tightly coupled to radiation
via electroweak interactions. This tight coupling prevented matter to cluster significantly at early times since
the radiation pressure acted as a counter force to the gravitational force.
As the Universe expands its mean energy density decreases, which eventually lead
to a freeze-out of these electroweak interactions. This freeze-out of the interactions (non-instantaneous in space and time)
happened roughly 380,000 years after the
Big Bang, eventually enabling the matter to cluster.
This epoch, usually called decoupling or recombination, marks the beginning of cosmological structure formation
with initially smooth matter density fluctuations.

The observed large-scale structure of the Universe
is mainly the result of the gravitational instability. 
To study the structure formation,
it is usually assumed that the CDM component behaves as a Newtonian pressureless and
curl-free fluid (so-called {\it cosmological dust}). 
Such a fluid is governed by the Euler--Poisson equations, together with 
the Friedmann equation, where the latter describes the background evolution of the expanding Universe.
Since that background evolution is by definition exactly homogeneous and isotropic, it is {parametrized}
by a single function (of cosmic time $t$): the cosmic scale factor $a(t)$.

{Analytical models for cosmological structure formation} can be formulated in either the Eulerian or Lagrangian frame
of reference (for some reviews, see e.g.~Bernardeau {\it et al.}~2002; Bernardeau 2013).
The class of analytical models of the latter is dubbed the Lagrangian perturbation theory (LPT)
(Buchert \& Goetz 1987; Buchert 1989; Buchert 1992; Bouchet {\it et
  al.}~1992; Bildhauer {\it et al.}~1992, Bouchet {\it et al.}~1995; Catelan 1995; Ehlers \& Buchert 1997).
{When using} this technique, the only dynamical variable 
 is the displacement, which represents the
gravitationally induced deviation of the particle trajectory field from the
homogeneous background evolution of the Universe.
To solve the Euler--Poisson equations in Lagrangian space, one usually requires some power series Ansatz
 for the displacement (for more information see also below in the Introduction).

The LPT is widely applied in cosmology.
Buchert (1992) 
showed that the Zel'dovich approximation (Zel'dovich 1970) is actually an instance of the first-order LPT (see also Novikov 1970). 
{Simply and elegantly, it} quite well describes the gravitational evolution of CDM inhomogeneities, as long as the trajectories of fluid particles do not cross.
 A Lagrangian approach has been also used in the so-called cosmological 
   reconstruction problem in an expanding Universe (Frisch {\it et al.}~2002;
 Brenier {\it et al.}~2003), where it is shown that, in so far as the dynamics
 are {governed} by the Euler--Poisson equations, the knowledge of both the
 present highly non-uniform distribution of mass and of its primordial
 quasi-uniform distribution, uniquely determines the inverse Lagrangian map,
 defined as the transformation from present Eulerian positions to
 their respective initial positions.
The LPT is also an important tool in numerical investigations. In  recent years,  second-order LPT  (2LPT)  has been successfully used to
set up initial conditions for numerical $N$-body
simulations (see e.g.~Crocce, Pueblas \& Scoccimarro~2006).
{The so-called biasing model}, whose task is to find a relationship between the visible matter and the dark matter distribution,
yield good predictions to cosmological observations when formulated in
Lagrangian space (Mo \& White 1996; Matsubara 2008b). 
Also, semi-analytical approaches to LPT deliver statistical estimators, such
as the matter power spectrum and bispectrum, which compare
favourably with results from $N$-body simulations (e.g., Matsubara
2008a; Rampf \& Wong 2012; Vlah, Seljak \& Baldauf 2014).  These
approaches make use of the LPT up to the fourth order (Rampf \&
Buchert 2012; Rampf 2012). Most of the LPT calculations are
performed for a flat CDM universe with vanishing cosmological constant,
i.e., for an Einstein--de Sitter (EdS) universe, and only few investigations, up to {the} third order, are known for a $\Lambda$CDM Universe (see, e.g., Matsubara 1995; Lee 2014).
The LPT has also been generalised to General Relativity, for a non-perturbative formulation see Buchert \& Ostermann (2012) and
Buchert, Nayet \& Wiegand (2013). For a general relativistic treatment within the  $\Lambda$CDM model see
 Rampf \& Rigopoulos (2013), Rampf (2013), Rampf \& Wiegand (2014).

Although LPT is widely applied in cosmology, very little is known about its
convergence properties, 
  which requires some knowledge about terms of arbitrarily
high order.  There are of course some exceptions, for example,
Sahni \& Shandarin (1996) investigated the case of an initial ``top-hat
underdensity'' that is initially discontinuous, and found that low-order
perturbation worked better than a higher-order one,
 which they regarded as a
possible evidence for ``semiconvergence'' of the perturbation series (see also
Moutarde {\it et al.}~1991).
Recently, a novel Lagrangian-coordinates  approach has been used to
show that particle trajectories for an EdS universe
are time analytical until a finite time (Zheligovsky \& Frisch 2014). 
Their approach is based on a little-known
Lagrangian formulation of ideal fluid flow derived by Cauchy in 1815
(Frisch \& Villone 2014).  In this paper we show that the approach of Zheligovsky
\& Frisch (2014) can be extended to a $\Lambda$CDM Universe (and even
beyond, see below), derive novel recursion relations
and prove time analyticity in the cosmic scale factor time, here used both
as a time variable and as an expansion parameter.

In the LPT literature, analytical results for $\Lambda$CDM usually 
employ an expansion in powers of
small displacements, which amounts to performing a (Taylor) expansion in powers of
the initial peculiar gravitational potential, $\varphi_g^{\rm (init)} \sim 10^{-5}$.
By this technique one seeks (reasonably well) \textit{approximate} solutions of the non-linear Euler--Poisson equations.
As a consequence of such an approximation technique, one is forced to solve
at each order in perturbation theory a second-order partial differential equation 
for the time coefficient of the displacement (the fastest growing solutions of these time coefficients are usually denoted with $D(t)$, $E(t)$, etc.). Because of that it is impossible to obtain recursion relations for $\Lambda$CDM by the use of such a ``small displacement'' expansion.
In this paper, by contrast,
 we perform \emph{a time-Taylor expansion} and seek for \textit{exact} analytic solutions for the fully non-linear Euler--Poisson equations in Lagrangian space.
As a consequence of our expansion scheme,
the displacement is represented in terms of a power series in the cosmic scale factor (even for a $\Lambda$CDM Universe!), and there is thus no need to explicitly solve partial differential equations for higher-order time coefficients.

This paper is organised as follows. In Section~\ref{sec:EPS} we
present various forms of  the
Euler--Poisson equations in a Eulerian-coordinate system and we show
that regularity of the solution at short times requires certain
{\it slaving constraints} on the initial conditions. In
Section~\ref{sec:Lag} we turn to the Euler--Poisson equations in Lagrangian
coordinates, which take a particularly simple form when using the cosmic scale factor as the time parameter. 
Recursion relations  are derived in Section~\ref{sec:rec}, from which
time-analyticity is derived in Section~\ref{sec:anal}. Further results
 related to time-analyticity are discussed in
Section~\ref{s:further}. This includes the dependence of the
time of guaranteed analyticity on the cosmological constant $\Lambda$
(Section~\ref{ss:obtainingtoflambda}), the absence of shell-crossing
during the time interval where analyticity can be proved
(Section~\ref{ss:shellcrossing}), the analyticity in the
linear growth time variable $D$ (Section~\ref{ss:D}) and the
persistence of analyticity when curvature effects are included,
as well as the problem that arises with radiation effects
(Section~\ref{ss:CR}).
Finally, in Section~\ref{sec:conc}, we summarise our main findings and
highlight various challenging open problems, such as using analyticity
 to develop a semi-Lagrangian numerical approach to the cosmological reconstruction.


\section{The various forms of the Euler--Poisson equations and the slaving conditions}
\label{sec:EPS}

A flat, matter dominated  Universe may be studied as a Newtonian and collisionless fluid whose governing equations are the Euler--Poisson equations.
For a flat Universe with cosmological constant, the Euler--Poisson equations are  (Peebles 1980) 
\begin{subequations}
 \begin{align}
   \label{eq:Eulermn}
&  \partial_t \tilde{\fett{U}} + \left( \tilde{\fett{U}} \cdot \nabr \right) \tilde{\fett{U}} =  - \nabr {\tilde\phi}_{g} \,, \\
& \label{eq:massmn}  \partial_t \tilde\varrho +\nabr \cdot \left( \tilde\varrho \tilde{\fett{U}}\right) =0 \,, \\
&  \label{eq:lambdapoisson} \nabr^2  \tilde\phi_g = 4\pi G \tilde\varrho - 3\Lambda \,.
\end{align}
\end{subequations}
Here, ${\fett{r}}$ is the proper space coordinate, $t$ the cosmic time,
$\tilde{\fett{U}}$ the velocity  of the fluid, its density $\tilde\varrho$; the
gravitational potential is denoted by $\tilde\phi_g$.
As to the cosmological constant, it is here denoted by $3\Lambda$.
Observe that, as long as we use the cosmic time as independent time variable,
the various dependent variables are surmounted by a tilde, 
which will be dropped when we change the time variable from cosmic time to 
cosmic scale factor $a$.
As usual, we decompose the mass density, fluid velocity and gravitational field respectively in a sum of two terms, one describing the effect of the uniform background expansion, the other the fluctuations against its background,
\begin{align} \label{decomp}
  \tilde\varrho = 
   \tilde{\bar\varrho}(t) \left[ 1 + \tilde\delta \right] \,, \qquad
  \tilde{\fett{U}} = \frac{\dot{a}}{a} {\fett{r}} + a \tilde{\fett{u}} \,, \qquad    \tilde{\phi}_g  = \tilde{\bar\phi}_g + \tilde{\varphi}_g \,.
\end{align}
Here, $\tilde \delta$ is the density contrast and $a \tilde{\fett{u}}$ the proper peculiar velocity.
Furthermore,
 $a(t)$ is the cosmic scale factor which {parametrizes} the global background expansion
  governed by the (first) Friedmann equation
\be \label{eq:Friedmann}
  \left( \frac{\dot{a}}{a}\right)^2 =  a^{-3} + \Lambda \,,
\ee
where we have used the fact that the background mass density is
$\tilde{\bar\varrho}(t)\! \sim \!a^{-3}(t)$, and we have set $8\pi G \bar\varrho_0 /3 =1$ for convenience.
The solution for the cosmic scale factor is easily obtained from the Friedmann
equation, namely  
\be
a(t) = \Lambda^{-1/3} [\sinh (3\sqrt{\Lambda}t/2)]^{2/3}\,,
\ee
which is proportional to $t^{2/3}$ for small $t$ (see e.g., Kofman, Gnedin \& Bahcall 1993).

After performing the aforementioned decomposition, derived in
Appendix~\ref{b:Nedsequations}, we obtain the Euler--Poisson equations
for  the  (comoving) peculiar velocity $\tilde{\fett{u}}$
\begin{subequations}
 \begin{align}
 \label{eq:Euler}
& \partial_t \tilde{\fett{u}} + \left( \tilde{\fett{u}} \cdot \nabx \right) \tilde{\fett{u}} = - 2 \frac{\dot{ a}}{ a} \tilde{\fett{u}}  - \frac{1}{{a}^2} \nabx \tilde \varphi_g \,, \\
 \label{eq:mass}
& \partial_t \tilde\delta +\nabx \cdot \left[ (1+\tilde\delta) \tilde{\fett{u}}\right] =0 \,, \\
 \label{eq:Pois}
& \nabx^2  \tilde{\varphi}_g = \frac{3}{2a} \tilde{\delta} \,.
 \end{align}
\end{subequations}
Note that the peculiar Euler--Poisson equations depend on the comoving coordinate $\fett{x} \equiv \fett{r}/a$.
Although these equations are indeed valid for a fluid model with cosmological
constant, the latter does not explicitly appear in the peculiar approach, but
only  implicitly through Friedmann's  background evolution equation
\rf{eq:Friedmann}.

Before considering the Euler--Poisson system~(\ref{eq:Euler}--\ref{eq:Pois}) in Lagrangian space, let us briefly discuss the linearised system in Eulerian space. Formally linearising around the steady state
$\tilde{\fett{u}}=\fett{0}$, $\tilde\delta=0$, we obtain the single differential equation for the density contrast,
\be \label{eq:evoL}
  \ddot{\tilde{\delta}} = -2 \frac{\dot a}{a} \dot{\tilde\delta} + \frac{3}{2a^3} \tilde\delta \,.
\ee
The solution of~\rf{eq:evoL} is most easily obtained by changing the time
variable from cosmic time $t$ to the cosmic scale factor $a$, here called $a$-time.
We thus set $\tilde\delta(t) = \delta(a(t))$. We then have $\partial_t = \dot a \partial_a$. The above differential equation is then
\be \label{eq:linEuler}
 (1+\Lambda a^3) \left[ \partial_{aa} \delta + \frac{3}{2a}  \partial_a \delta \right] + \frac{3}{2} \Lambda a^2 \partial_a \delta - \frac{3}{2a^2} \delta = 0 \,,
\ee
where we have used the Friedmann equation\,(\ref{eq:Friedmann}).
This equation has two solutions. One is called the decaying
mode; it behaves as $a^{-3/2}$ for small
$a$ and thus blows up when $a\rightarrow 0$,  thereby invalidating the
linearisation. The other one, the growing mode, here called the linear
growth function, is taken to be
 $D(a) = a \sqrt{1+ \Lambda a^3} \, {}_2F_1\!\left( 3/2, 5/6, {11}/{6}, -
\Lambda a^3 \right)$, where ${}_2F_1$ is the Gauss hypergeometric
function (Demianski {\it et al.}~2005; Enqvist \& Rigopoulos 2010; Hamber \& Toriumi 2010). 
This solution is analytic around
$a=0$, has the small $a$ expansion  $D(a) = a - (2/11)\,\Lambda a^4 +
\text{h.o.t.}$
and thus essentially reduces to  the cosmic scale factor at short times. This is the
physically appropriate solution, in which density fluctuations are growing linearly with $a$.

However, when studying the well-posedness of the non-linear
Euler--Poisson system (\ref{eq:Euler}--\ref{eq:Pois}) and, subsequently, when looking for analytic solutions
in Lagrangian coordinates, a linear {approach ceases to be appropriate}.
 The latter just gives us a signal that
the appropriate time variable might be proportional to $t^{2/3}$ or to $a$ or
 $D$ at short times. The non-linear theory can be developed in all three
such time variables but, for our purposes, it takes its simplest form when using
the $a$-time (and not just at short ones). Indeed the Euler--Poisson system
(\ref{eq:Euler}--\ref{eq:Pois}) can be rewritten in the following form:
\begin{subequations}
\begin{align}
  \label{eq:Eulera}
&\!\left(  1+ \Lambda a^3 \right) \left[  \partial_a \fett{v} + \left( \fett{v} \cdot \nabx \right) \fett{v} \right] = -
\frac{3}{2a} \left( \fett{v} + \nabx \varphi_g \right) - 3 \Lambda a^2 \fett{v} \,, \\
   \label{eq:massa}
&\partial_a \delta +\nabx \cdot \left[ \left( 1+ \delta \right) \fett{v} \right] = 0  \,, \\
   \label{eq:Poisa}
& \nabx^2 \varphi_g = \frac \delta a \,,
\end{align}
\end{subequations}
where we used a new velocity variable $\fett{v}$, the $a$-derivative of comoving
particle positions, related to the old velocity  by
\be \label{eq:defv}
\tilde{\fett{u}}(\fett{x},t) = \dot a \fett{v}(\fett{x},a(t))\,,
\ee
and we have set $\tilde\varphi_g(\fett{x},t) = (3/2) \varphi_g(\fett{x},a(t))$.
Note that, with respect to $\Lambda$, the situation in (\ref{eq:Eulera}--\ref{eq:Poisa}) is the opposite of what we had with~\mbox{(\ref{eq:Euler}--\ref{eq:Pois}):}
$\Lambda$ appears now explicitly, but the Friedmann equation can now be omitted as
long as we do not want to revert to the cosmic time variable.

An important feature of (\ref{eq:Eulera}--\ref{eq:Poisa}) is the presence
of cosmic scale factors in the denominators of the r.h.s. of \eqref{eq:Eulera}
and \eqref{eq:Poisa}. It  indicates that the solution will be singular
at $a=0$ unless the corresponding numerators also vanish, that is, we have to
satisfy the two following \textit{slaving conditions} (Brenier~1987) at
$a=0$ (denoted by the superscript (init)):
\be \label{slaving} \delta ^{\rm (init)} = 0, \qquad \fett{v}^{\rm (init)} =
-\nabx \varphi_g^{\rm (init)}\,.\ee The second slaving condition immediately
implies the curl-free character of the initial velocity, which persists
by~(\ref{eq:Eulera})
in Eulerian (but not in Lagrangian) coordinates.
Traditionally, this irrotationality is often derived by linearising the
Euler--Poisson equations and showing that the rotational component of the
velocity decays in  time (see e.g., Bernardeau {\it et al.}~2002). This linearisation would be justified if the velocity $\fett{v}$ would be small
at short times as indeed the density contrast $\delta$ is. However, the
velocity $\fett{v}$ and the gravitational force $\nabx \varphi_g$ do
not vanish as $a\to 0$. Thus, a linearisation-based argument is questionable.

The cosmic scale factor $a$, being dimensionless, is conveniently normalised
to unity at the present-time epoch. At decoupling it then has a value of about
$10^{-3}$. Here, we let  the cosmic scale factor start at the value zero, while using
the Euler--Poisson model with slaving. As a consequence,
the whole primordial (pre-decoupling) cosmology is just reduced  to the
prescription of the slaving conditions \rf{slaving}. 
We shall come back to this issue in the Concluding Remarks~\ref{sec:conc}. 

\section{The Lagrangian formulation of $\Lambda$CDM} \label{sec:Lag}

Time-Taylor expansions can be
carried out in either Eulerian or in Lagrangian coordinates.
For such expansions to be convergent, {i.e., for
the radius of convergence not to vanish}, it is much preferable to work in Lagrangian
coordinates. Indeed, if the initial conditions have spatial
derivatives only up to some finite order, the Eulerian solutions will
have  time derivatives only up to the same order. The reason is that, when
one takes a time derivative in Eulerian coordinates, this
generates a space derivative, because of the $(\fett{v} \cdot \nabx )
\fett{v}$ term. Hence time-Taylor coefficients will not exist beyond
that order. This does not happen in Lagrangian coordinates, where a
limited
amount of smoothness in the initial data suffices to ensure
time-analyticity (Zheligovsky \& Frisch 2014). 
Even when the initial data are spatially analytic,
so that the solutions are time-analytic in both Eulerian and
Lagrangian coordinates, the Lagrangian radius of convergence, being
controlled by the largest strain in the initial data, is typically
much larger than the Eulerian one, which depends on the largest
initial velocity present (for more details on such matters, see
Podvigina {\it et al.}~2015). 

We now turn to the Lagrangian formulation of the Euler--Poisson equations
(\ref{eq:Eulera}--\ref{eq:Poisa}). For this we use the (direct) Lagrangian
map $\fett{q} \mapsto {\fett{x}}(\fett{q},a)$ from the initial ($a=0$)
position
$\fett{q}$ to the Eulerian position
${\fett{x}}$ at time $a$.
The velocity ${\fett{v}}$, as it is apparent from its definition~\eqref{eq:defv},
is simply the Lagrangian $a$-time derivative of the position;
more precisely, 
 $\fett{v}(\fett{x}(\fett{q},a),a) = \partial_a ^{\rm L}\fett{x}(\fett{q},a)$, 
where $\partial_a^{\rm L}$ is the Lagrangian
$a$-time derivative.
At initial time, $a=0$, the velocity is
\be
 \fett{v}^{(\rm init)}(\fett{q}) =  \fett{v}(\fett{x}(\fett{q},0),0)  \,, \label{eq:vinit}
\ee
which coincides with the Eulerian one.
We shall also use
the Jacobian matrix $\nabla_j ^{\rm L} x_i(\fett{q},a)$, where $\nabla_j ^{\rm
    L}$ denotes the Lagrangian space derivative with respect to $q_j$, and its
  determinant $J = \det \left(\nabla_j ^{\rm L} x_i\right)$, the
  Jacobian. Note that after shell-crossing, i.e., the first vanishing of $J$, the
  map ceases to have a unique inverse.

As an intermediate step to a fully Lagrangian formulation of the
Euler--Poisson equations, we rewrite the momentum equation
\eqref{eq:Eulera} and the Poisson equation \eqref{eq:Poisa} in a
mixed form: some terms are more naturally expressed in Lagrangian coordinates
and others in Eulerian coordinates; composition with the direct or
inverse  Lagrangian maps is here understood where appropriate:
\begin{subequations}
\begin{align}
  \label{eq:Eulerb}
& \!\left(  1+ \Lambda a^3 \right) \partial_{aa} ^{\rm L} \fett{x} = -
\frac{3}{2a} \left( \partial_{a} ^{\rm L} \fett{x} + \nabx \varphi_g \right) - 3 \Lambda a^2  \partial_{a} ^{\rm L} \fett{x}\,, \\
   \label{eq:massb}
& ( 1+ \delta )J =1 \,, \\
   \label{eq:Poisb}
& \nabx^2 \varphi_g = \frac \delta a \,.
\end{align}
\end{subequations}
Equation~\eqref{eq:massb} is a straightforward expression of mass
conservation in Lagrangian coordinates, which need not be derived from its
differential Eulerian counterpart. It is however invalid after shell-crossing.
Due to the appearance of several velocities (multi-streaming), there are several branches of the inverse Lagrangian map. In that case
\rf{eq:massb} must be replaced by
\be
( 1+ \delta ) = \sum_n \frac{1}{|J_n|}\,,
\label{sumoverabsvalues}
\ee
where $n$ labels the various Lagrangian locations having the same Eulerian
location (Buchert~1995). 
It is likely that each branch of the
multi-streaming system still satisfies the Euler--Poisson equations with the
density given by \rf{sumoverabsvalues}, but this requires further investigation.

The derivation of the full Lagrangian formulation of the Euler--Poisson
equations has two parts. The first part is almost identical to Cauchy's 1815
derivation of the Cauchy invariants equation  for
incompressible fluids  (see Frisch \& Villone 2014). Since it is already given
in Zheligovsky \& Frisch (2014)  for the case of an EdS
universe, we shall just recall the ideas briefly 
 and give the final
Cauchy equation. We observe that in the r.h.s. of \eqref{eq:Eulerb}, because
of the curl-free character of the velocity  $\partial_{a} ^{\rm L}
\fett{x}$, all the terms are Eulerian gradients. Multiplication by the
Jacobian matrix changes these Eulerian gradients to Lagrangian
gradients. These are then made to vanish by taking a Lagrangian curl. After
this, a quadratically non-linear equation  is obtained for the Lagrangian map,
which involves first and second time derivatives. Cauchy observed that this
equation can be exactly integrated in time. This ``miracle'' has been
interpreted much later as  stemming
from the relabeling invariance of the variational formulation of the Euler
equation and use of Noether's theorem (Frisch \& Villone 2014, Section~5.2).
 The resulting Cauchy
invariants equations have normally the initial vorticity in its r.h.s but, in
the present case, the vorticity vanishes and the equations take the following
form:
\be \label{eq:main2}
  \varepsilon_{ijk}  \nabL{j} x_l\, \partial_a^{\rm L} \nabL{k}  x_l = 0 \,, \qquad \qquad (i=1,2,3)
\ee
where $\varepsilon_{ijk}$ is the fundamental antisymmetric tensor, 
 and summation over repeated indices is assumed.
The l.h.s.~of \eqref{eq:main2} is a Lagrangian curl.

The second part
makes use of all three equations \eqref{eq:Eulerb}-\eqref{eq:Poisb}, and gives
a third scalar equation:
\be
\label{eq:main}
\varepsilon_{ikl} \varepsilon_{jmn} \nabla_m^{\rm L} x_k \nabla_n^{\rm L} x_l \left[ \left( 1+\Lambda a^3 \right) \partial^{\rm L}_{aa} + 3 \Lambda a^2 \partial^{\rm L}_{a} + \frac{3}{2a} \partial^{\rm L}_{a} \right] \nabla_j^{\rm L} x_i
    = \frac{3}{a^2} \left( J-1 \right) \,.
\ee
This is done by taking the Eulerian divergence of
\eqref{eq:Eulerb} to obtain the Eulerian Laplacian of the
gravitational
potential, which is then substituted in \eqref{eq:Poisb}. The density
contrast $\delta$ is expressed in terms of the Jacobian, using \eqref{eq:massb}
and also
substituted in the r.h.s. of \eqref{eq:Poisb}. Finally, the Eulerian
divergence is expressed in terms of Lagrangian space derivatives and
of the inverse Jacobian matrix. The latter is given in terms
of the Jacobian matrix using the following  identity:
$\nabla_{x_i}
q_j = \varepsilon_{ikl}
\varepsilon_{jmn} \nabL{m} x_k \nabL{n}  x_l/(2J)$, which follows from the
observation that the Jacobian {matrix} of the inverse Lagrangian map is the matrix inverse of
the Jacobian of the direct map (see, e.g., Buchert \& Goetz 1987).
This derivation fails if
the Jacobian $J$ is not everywhere strictly positive in  the space and time
domain under consideration. In the latter case, the Lagrangian formulation
\rf{eq:main2}-\rf{eq:main} becomes invalid. We shall come back to this issue in Section~\ref{ss:shellcrossing}.

\section{The formal Taylor expansion and the recursion relations}\label{sec:rec}

In the LPT literature, analytical derivations of results
 for $\Lambda$CDM usually rely on
 an expansion in powers of small displacements, which amounts to performing
a power-series expansion in the initial peculiar gravitational potential
(see e.g., Matsubara~1995; Bernardeau {\it et al.}~2002).
In this paper we follow a different
 approach and perform an expansion in powers of $a$-time.
Note that for the case of an EdS universe, one can obtain recursion relations
 for the displacement by both expansion techniques, and the resulting recursion relations are formally identical (Rampf 2012). For a $\Lambda$CDM Universe,
however, explicit results for the displacement will generally differ
depending on what expansion scheme is used.

We thus seek a solution to the Lagrangian equations~(\ref{eq:main2})--(\ref{eq:main}) in the form of a power series in the $a$-time (i.e., a Taylor series) 
for the displacement
 $\fett{\xi} \equiv \fett{x} - \fett{q}$,
\be \label{ansatzDispl}
  \fett{\xi}(\fett{q},a) = \sum_{s=1}^\infty \fett{\xi}^{(s)}(\fett{q})\, {a}^s \,.
\ee
At first order,  we have
\be
  \fett{\xi}^{(1)}(\fett{q}) = \fett{v}^{(\rm init)} (\fett{q})\,, \label{eq:firstorder}
\ee
as follows from $\fett{v} = \partial_a^{\rm L} \fett{x}$ and~(\ref{eq:vinit}).
The convergence of the series \rf{ansatzDispl} will be examined in the next section.
The Jacobian of the Lagrangian map $J= \det(I+\nab ^{\rm
  L}\fett{\xi})$, where $I$ is the identity matrix, can be written as
a sum of four terms. After substituting the expansion \eqref
{ansatzDispl}, the Jacobian becomes
\be \label{jacobian}
   J = 1 + \sum_{s=1}^\infty  \mu_1^{(s)} a^s
    + \sum_{n_1+n_2 = s}^\infty \mu_2^{(n_1,n_2)} a^s + \sum_{n_1+n_2+n_3 = s}^\infty \mu_{3}^{(n_1,n_2,n_3)} a^s \,,
\ee
where the sums are restricted to values of $n_1$, $n_2$ and $n_3$ that
are strictly positive and where the various quantities $\mu_1$, $\mu_2$ and $\mu_3$ are
expressions linear, quadratic and cubic, respectively, in the Lagrangian gradients of
the Taylor coefficients, {given by}
\begin{align}
&   \mu_1^{(n_1)} = \fett{\nabla}^{\rm L} \cdot \fett{\xi}^{(n_1)} \,, \\
&   \mu_2^{(n_1,n_2)} = \frac 1 2 \left( \nabla_i^{\rm L} \xi_{i}^{(n_1)}  \nabla_j^{\rm L} \xi_{j}^{(n_2)} -  \nabla_j^{\rm L} \xi_{i}^{(n_1)}  \nabla_i^{\rm L} \xi_{j}^{(n_2)}  \right) \,, \\
&   \mu_3^{(n_1,n_2,n_3)} = \frac 1 6 \varepsilon_{ikl} \varepsilon_{jmn} \nabL{m} \xi_k^{(n_1)} \, \nabL{n} \xi_l^{(n_2)} \, \nabL{j} \xi_i^{(n_3)} \,.
\end{align}

Substituting the expansion~(\ref{ansatzDispl}) into (\ref{eq:main})
and identifying the coefficients of the various powers of $a$, we find,
for $s>0$,
\begin{align}
  & \left(s+ \frac 3 2 \right) (s-1)\, \mu_1^{(s)} =
 \sum_{n_1+n_2 = s}  \left\{ \left[ \frac 3 2 - 2 n_2 \left( n_2+ \frac 1 2 \right) \right] \mu_2^{(n_1,n_2)}  -2 \Lambda \left( n_2 -3 \right) \left( n_2  -1 \right)
   \mu_2^{(n_1,n_2-3)} \right\} \nonumber \\
 &\qquad
   -\Lambda \left( s-3 \right) \left( s-1 \right) \mu_1^{(s-3)} + \sum_{n_1+n_2+n_3 = s} \left\{ \left[ \frac 3 2 - 3 n_3 \left(n_3 + \frac 1 2 \right)  \right] \mu_3^{(n_1,n_2,n_3)} - 3 \Lambda  \left( n_3-3 \right) \left( n_3 -1 \right)  \mu_3^{(n_1,n_2,n_3-3)}  \right\}  \,.\label{beggingforalabel}
\end{align}
Here, by construction, 
all the $\mu$ coefficients vanish if one or several of their upper indices are
zero or negative.
We symmetrise the r.h.s.~and obtain a sequence of relations for $s \geq 1$:
\begin{align}
  \fett{\nabla}^{\rm L} \cdot \fett{\xi}^{(s)} \equiv  \mu_1^{(s)} &=
  \fett{\nabla}^{\rm L} \cdot  \fett{v}^{(\rm init)} \delta_1^s  -\Lambda \frac{s-3}{s+ 3/2} \mu_1^{(s-3)}  \nonumber \\
 & +   \sum_{0 < n < s} \left\{ \frac{( 3 - s)/ 2 - n^2 - (s-n)^2}{(s+ 3/2) (s-1)} \mu_2^{(n,s-n)}  + \Lambda \frac{4s -n^2 - (s-n)^2 -6}{(s+ 3/2) (s-1)}
   \mu_2^{(n,s-n-3)} \right\} \nonumber \\
 & + \sum_{n_1+n_2+n_3 = s} \left\{ \frac{(3 -s) /2 - n_1^2 -n_2^2 -n_3^2}{(s+ 3/2) (s-1)} \mu_3^{(n_1,n_2,n_3)} - \Lambda  \frac{n_1^2 +n_2^2 + n_3^2 -4s + 9}{(s+ 3/2) (s-1)}  \mu_3^{(n_1,n_2,n_3-3)}  \right\}  \,, \label{eq:recLongitudinal}
\end{align}
where $\delta_1^s$ is the Kronecker symbol.
Similarly, we obtain from the Cauchy invariants equation~(\ref{eq:main2}), for $s \geq 1$
\be \label{eq:recTransverse}
  \fett{\nabla}^{\rm L} \!\times \fett{\xi}^{(s)} \equiv \fett{T}^{(s)} = \frac 1 2 \sum_{\substack{\,1 \leq k \leq 3,\\  0 < n <s }}  \frac{s-2n}{s} \nab^{\rm L} \xi_k^{(n)} \times \nab^{\rm L} \xi_k^{(s-n)}   \,.
\ee
As we see, \eqref{eq:recLongitudinal} and \eqref{eq:recTransverse} give
  the Lagrangian divergence and curl of the $a$-time-Taylor
  coefficient of order $s$ in terms of lesser-order ones.
We observe that the first explicit occurence of
$\Lambda$ in the recursion relations {is}
at the fourth order, which is the same order at which the linear growth $D(a)$,
at short $a$-times,
differs from its EdS value: $D(a) = a -(2/11)\Lambda a^4+ O(a^7)$.

Eqs.~\eqref{eq:recLongitudinal} and \eqref{eq:recTransverse}
give a Helmholtz--Hodge decomposition of $\fett{\xi}^{(s)}$ into
curl and divergence, from which it can be retrieved in a standard way 
(cf.~e.g.~Arfken \& Weber 2005), namely
\be
 \fett{\xi}^{(s)} = \nab^{-2} \left( \nab^{\rm L} \mu_1^{(s)} - \nab^{\rm L} \times \fett{T}^{(s)} \right) \,,  \label{eq:HH}
\ee
where $\nab^{-2}$ is the inverse Laplacian in Lagrangian coordinates 
({with the boundary conditions taken into account}), and
 $\mu_1^{(s)}$ and  $\fett{T}^{(s)}$ denote the r.h.s.'s of Eq.\,(\ref{eq:recLongitudinal})
and Eq.\,(\ref{eq:recTransverse}), respectively.

Using~\eqref{eq:HH} and
considering the first-order Lagrangian derivatives of  $\xi_\nu^{(s)}$ we
can combine the two recursion relations \eqref{eq:recLongitudinal} and \eqref{eq:recTransverse} into a single recursion relation for the gradient tensors $\nabL{\mu} \xi_\nu^{(s)}$ of the time-Taylor coefficients, which reads
\begin{align}
 \nabL{\mu} \xi_\nu^{(s)} &= \nabL{\mu} {v}_\nu^{\rm (init)} \delta_1^s
-\Lambda \frac{s-3}{s+ 3/2}  {\cal C}_{\mu \nu} \mu_1^{(s-3)}
+ \sum_{\substack{\,1 \leq j \leq 3, \\ j\neq \nu }} {\cal C}_{\mu j}
\left( \sum_{\substack{\,1 \leq k \leq 3,\\  0 < n <s }}
\frac{2n-s}{s} \left(\nabL{\nu} \xi_k^{(n)}\right)
\nabL{j} \xi_k^{(s-n)} \right) \nonumber \\
   &+  {\cal C}_{\mu \nu} \superbig( \sum_{0 < n < s} \left\{ \frac{( 3 - s)/ 2 - n^2 - (s-n)^2}{(s+ 3/2) (s-1)} \mu_2^{(n,s-n)}  + \Lambda \frac{4s -n^2 - (s-n)^2 -6}{(s+ 3/2) (s-1)}
   \mu_2^{(n,s-n-3)} \right\} \nonumber \\
 &\qquad \qquad + \sum_{n_1+n_2+n_3 = s} \left\{ \frac{(3 -s) /2 - n_1^2 -n_2^2 -n_3^2}{(s+ 3/2) (s-1)} \mu_3^{(n_1,n_2,n_3)} + \Lambda  \frac{4s - 9-n_1^2 -n_2^2 - n_3^2}{(s+ 3/2) (s-1)}  \mu_3^{(n_1,n_2,n_3-3)}  \right\} \superbig) \,, \label{eq:nabRec}
\end{align}
where ${\cal C}_{ij} \equiv \nab^{-2} \nabL{i} \nabL{j}$ is an
operator of the Calderon--Zygmund type (cf.~Zheligovsky \& Frisch 2014).
The recursion relations \rf{eq:nabRec} are explicit but look
somewhat lengthy.  Actually, they enjoy an 
important property which allows us to prove time-analyticity
in the next section: for $s>1$, it is easily checked that  all the coefficients 
involving $s,\,n_1,\,\ldots$ are bounded (rational) functions. 
{In those not involving $\Lambda$, the coefficients are bounded by unity and the other ones by $\Lambda>0$.}

\section{The time-analyticity of the Lagrangian solution}\label{sec:anal}

After having found explicit recursion relations for the time-Taylor series
of the displacement field, {we are naturally led to ask:} is
\be \label{ansatzDisplRepeated}
\fett{\xi}(\fett{q},a) = \sum_{s=1}^\infty \fett{\xi}^{(s)}(\fett{q})\, a^s
\ee
 a convergent series that thus
defines a  time-analytic function in the neighbourhood of the origin?

Let us give first some overall ideas {on} how the convergence results will be
established. 
Given that the recursion relations take their most compact form \rf{eq:nabRec} in terms of
gradients, it is simpler to first prove the convergence for the gradient series
of
\be \label{ansatzDisplTensor}
\nabL{\mu}{\xi}_\nu(\fett{q},a) = \sum_{s=1}^\infty \nabL{\mu}{\xi}_\nu^{(s)}(\fett{q})\,a^s \,. 
\ee
From the recursion relations \rf{eq:nabRec} for the gradients of the Taylor
coefficients, we shall be able to derive
polynomial recursion \textit{inequalities} for their norms. By reintroducing
the $a$-time and summing over all orders (i.e., by using a generating
function), 
we shall obtain a single inequality for an $a$-dependent cubic polynomial. The study
of the evolution of its roots will give us the required bounds on the norms of the
time-Taylor coefficients of the gradient series, from which the convergence of the
series~(\ref{ansatzDisplTensor}) and thus time-analyticity  are established. 
The analyticity of the Lagrangian
map and of the displacement field are then a consequence.

Let us proceed now with the details. 
First, we observe that the r.h.s.'s of the recursion relations \rf{eq:nabRec}
for the gradient tensors are themselves mostly polynomials (of degree
not exceeding three) of the lesser-order gradient tensors. We write
``mostly'' because there are also the Calderon--Zygmund operators,
${\cal C}_{ij}$. In the present case, these operators stem from the nonlocal
nature of the gravitational interaction. 
 Technically speaking, these are
pseudo-differential operators of degree zero (because an inverse
Laplacian is compensated by two space derivatives). Essentially, such
operators do not change the degree of differentiability of the
functions they are applied to. But this is true only when
using a suitable function space in which the Calderon--Zygmund
operators are bounded.

For our purposes the function space in which such matters are simplest
is the space $\ell_1$
of spatially periodic functions, say, of periodicity $2\pi$ in
all three space variables and such that their Fourier series is
absolutely summable. Concretely, 
let a periodic function $\mathbf{f}(\fett{q})$
(of scalar, vector or tensor type)
be expanded in a Fourier series:
  \be \label{ell1}
  \mathbf{f}(\fett{q}) = \sum_{\fett{k}} \hat{\mathbf f}_{\fett{k}} \,{\rm e}^{\ii \fett{k} \cdot \fett{q}} \,,
\ee
where the summation is over all triplets $\fett{k}$ of signed integers.
A periodic function is said to be in $\ell_1$, if the 
 norm
\be
  \| \mathbf{f} \| \equiv \sum_{\fett{k}} |\hat{\mathbf{f}}_{\fett{k}}|
\ee
is finite (for vector and tensor quantities, the modulus is defined as
the maximum over all indices of the modulus of the various components). Since, for any $i$ and $j$, one has $|k_ik_j/\fett{k}^2|
\le 1$, the Calderon--Zygmund operators are obviously bounded by unity in this
space. It is also elementary to show that this space enjoys the 
algebra property:
\be
\| \mathbf{f}\,\mathbf{g}\| \le \| \mathbf{f}\| \| \mathbf{g}\|,
\ee
for any pair of functions $\mathbf{f}$ and $\mathbf{g}$ of finite $\ell_1$ norm.
Other function spaces with similar properties may be
more realistic for cosmological applications in so far as periodicity
is not required, such as H\"older spaces. For this we refer the reader
to Section 2.5 of Zheligovsky \& Frisch (2014).

Given the structure of the recursion relations \rf{eq:nabRec}  and
using the boundedness of the Calderon--Zygmund operators
 and the algebra properties of the  $\ell_1$ norm,
it is elementary to show that if $\nabLT \fett{v}^{(\rm init)}$ is in
the space $\ell_1$,  so will be the gradients of the Taylor coefficients
$\nabLT \fett{\xi}^{(s)}$. The assumption that the gradient of
the initial velocity has absolutely summable Fourier coefficients
is the key hypothesis of the present work.

Knowing that all the (Lagrangian) gradients of the Taylor coefficients are in
$\ell_1$ is not enough to ensure the convergence of the gradient Taylor
series.  For this, we need to obtain suitable bounds on the $\ell_1$ norms of
$\nabLT \fett{\xi}^{(s)}$, and then we need to show that these bounds imply
the convergence of the gradient time-Taylor series.  Here we work with the
time-Taylor series for the spatial gradient of the displacement and establish
its time-analyticity, from which follows the time-analyticity of the
displacement (and also, of course, of the full Lagrangian map). To obtain the
bounds on the $\ell_1$ norms, we shall mostly follow the approach presented in
Zheligovsky \& Frisch (2014). Of course, due to the presence of the
cosmological constant, the third-order polynomials contain additional terms,
and their study is more involved.

The first step is to bound the  various  $\| \nabLT \fett{\xi}^{(s)} \|$,
using the boundedness of the Calderon--Zygmund operators, the algebra property,
and the boundedness of the rational
coefficients, as explained at the end of
section~\ref{sec:rec}. We use~\rf{eq:nabRec} and thereby obtain, for any $s\ge  1$:
\begin{align} \label{eq:boundednorm}
  \left\| \nab^{\rm L} \fett{\xi}^{(s)}  \right\|  &\leq
    \left\| \nab^{\rm L} \fett{v}^{\rm (init)}  \right\| \delta_1^s 
    + 3 \Lambda \left\| \nab^{\rm L} \fett{\xi}^{(s-3)}  \right\|
  + 12 \sum_{i+j=s} \left\| \nab^{\rm L} \fett{\xi}^{(i)} \right\| \, \left\| \nab^{\rm L} \fett{\xi}^{(j)}  \right\|
  +6 \Lambda \sum_{i+j=s} \left\| \nab^{\rm L} \fett{\xi}^{(i)} \right\| \, \left\| \nab^{\rm L} \fett{\xi}^{(j-3)}  \right\|
   \nonumber \\
  &\qquad
 + 6 \sum_{i+j+k= s} \left\| \nab^{\rm L} \fett{\xi}^{(i)}  \right\| \, \left\| \nab^{\rm L} \fett{\xi}^{(j)}  \right\| \, \left\| \nab^{\rm L} \fett{\xi}^{(k)}  \right\| 
 + 6 \Lambda \sum_{i+j+k= s} \left\| \nab^{\rm L} \fett{\xi}^{(i)}  \right\| \, \left\| \nab^{\rm L} \fett{\xi}^{(j)}  \right\| \, \left\| \nab^{\rm L} \fett{\xi}^{(k-3)}  \right\|
   \,.
\end{align}

\begin{figure}
\includegraphics[width=\textwidth]{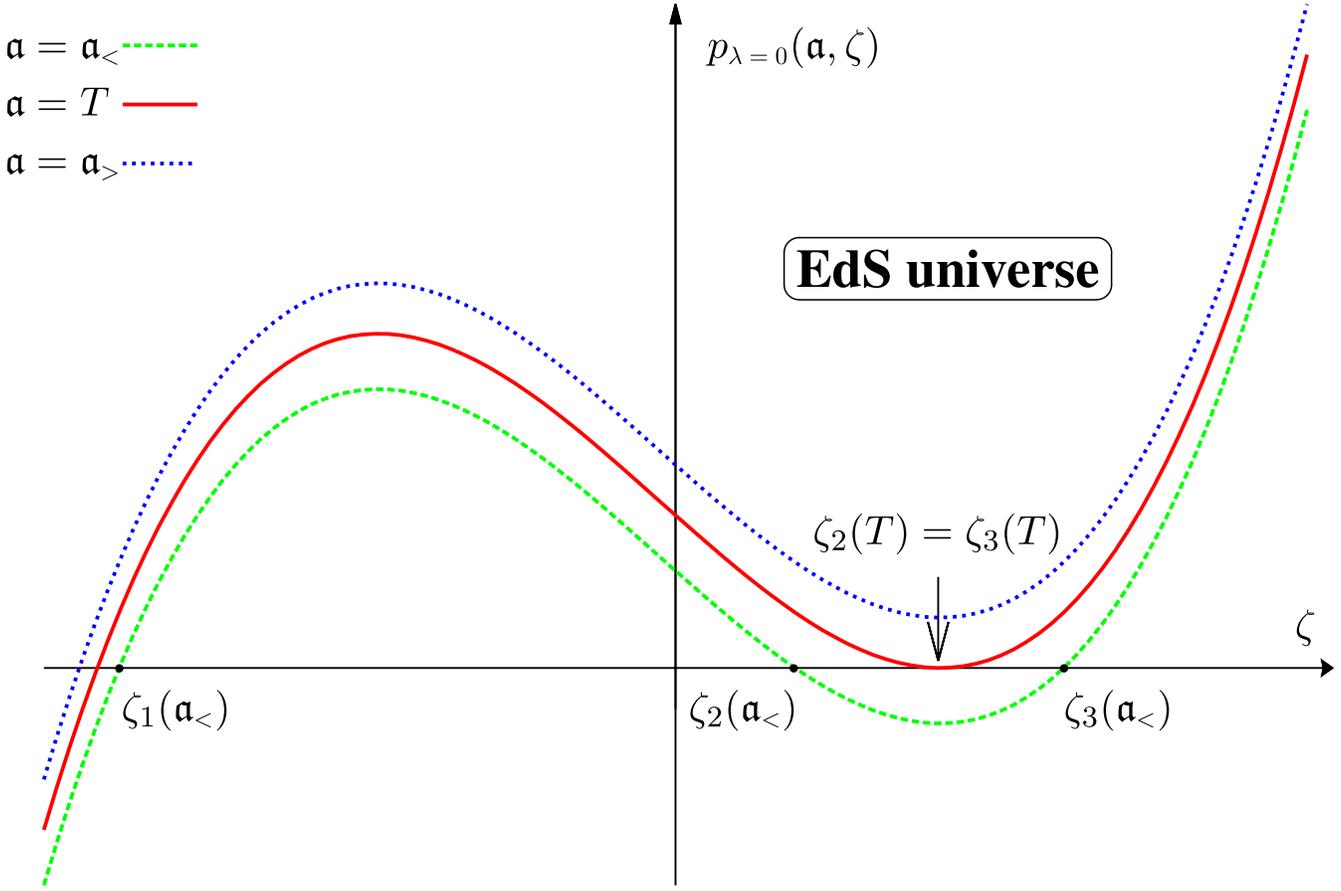}
\caption{Sketch of the 
polynomial~(\ref{pol}),  for $\lambda=0$, i.e., for the case of an EdS universe. Shown
are three values of the rescaled time  $\mathfrak{a}$ for which $0 < \mathfrak{a}_{\text{\tiny $<$}} < T(\text{\scriptsize $\lambda=0$}) < \mathfrak{a}_{\text{\tiny $>$}}$,
illustrating the behaviour of real roots of $p_0(\mathfrak{a},\zeta)$ (the roots are shown as the points of intersection of the graph of
$p_{0}(\mathfrak{a}_{\text{\tiny $<$}},\zeta)$ and the horizontal axis). On increasing $\mathfrak{a}$, the graph slides up as a rigid curve. As a result, the roots
$\zeta_1$ and $\zeta_3$ move to the left (i.e.~become smaller), whereas $\zeta_2$ moves to the right (i.e.~becomes larger). At the critical value
$T(0)$, when $\partial p_{0} / \partial \zeta = 0$ and the discriminant $\Delta$ vanishes, the two roots $\zeta_2$
and $\zeta_3$ collide and then disappear (with the emergence of a pair of complex conjugate roots).}
\label{fig:analEdS}
\end{figure}

\begin{figure}
\includegraphics[width=\textwidth]{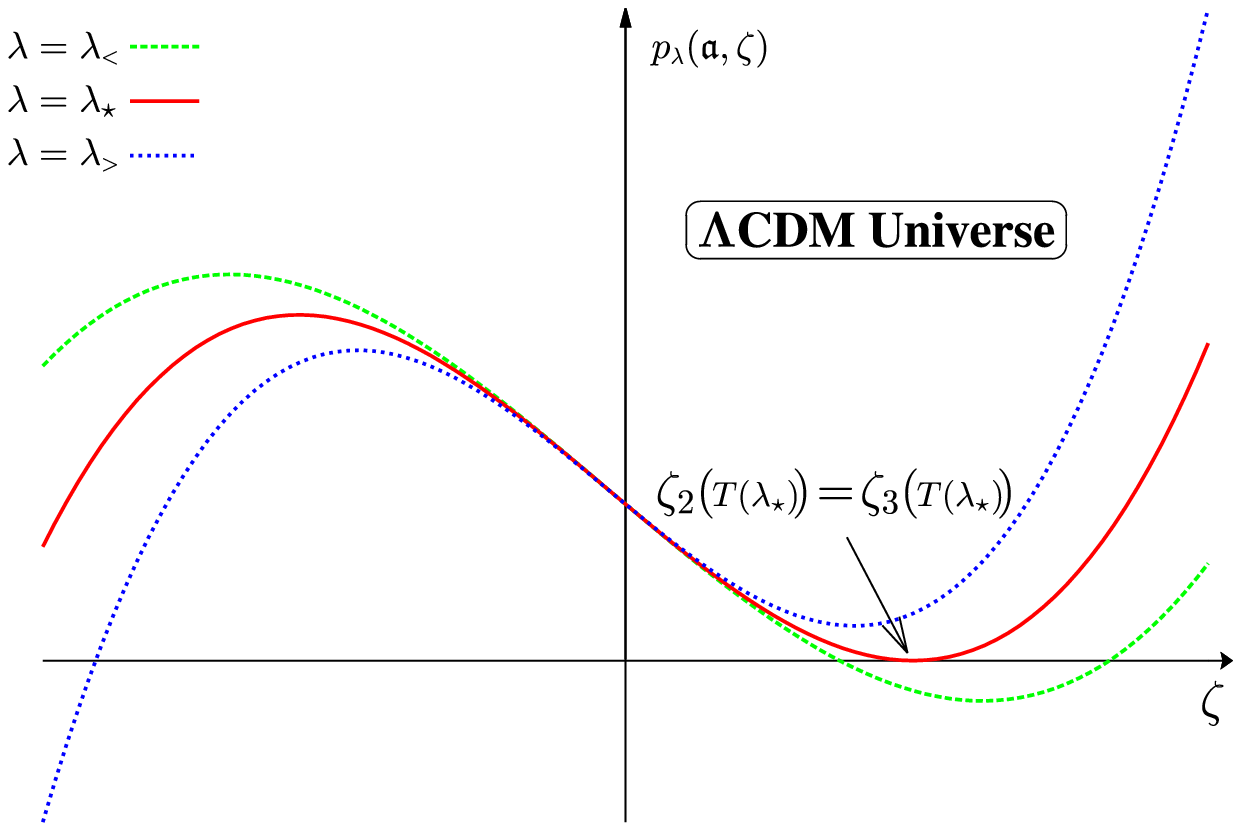}
\caption{
Sketch of the polynomial~(\ref{pol}) for any fixed rescaled time variable
 $\mathfrak{a}>0$ and $\lambda \neq 0$ (i.e., the $\Lambda$CDM case).
Three values of lambda are shown for
which  $\lambda_{\text{\tiny $<$}} < \lambda_\star < \lambda_{\text{\tiny $>$}}$.
The value $\lambda_\star$ denotes the critical value for which the cubic polynomial $p_\lambda$ develops a double root, i.e., where $\zeta_2(T(\lambda))=\zeta_3(T(\lambda))$.
In that case the fixed value of the rescaled time $\mathfrak{a}$ is precisely the critial value $T(\lambda_\star)$ for which the generating function $\zeta$ of the displacement is bounded.
On increasing $\lambda$  for any fixed rescaled time $\mathfrak{a} >0$, the graph gets shifted into the upper left direction.
}
\label{fig:analvaryomega}
\end{figure}

The second step is to introduce a generating function 
{using gradients of Taylor coefficients}, namely
\be \label{eq:generate} {\tilde \zeta}(a) \equiv \sum_{s=1}^\infty \left\| \nab^{\rm L}
\fett{\xi}^{(s)} \right\| a^s \,.
\ee
It is easily checked that this generating function is an upper bound for the
$\ell_1$ (absolutely summable Fourier series) norm  
of the gradient of the displacement field at time
$a$. Multiplying \rf{eq:boundednorm} by $a ^s$ and summing over $s$ from one
to infinity, we obtain the following  inequality for the generating function:
\be 
  {\tilde \zeta} \leq  a \left\| \nab^{\rm L} \fett{v}^{\rm (init)}  \right\|   + 12
  {\tilde \zeta} ^2
 + 6 {\tilde \zeta} ^3 +6 \Lambda a ^3 {\tilde \zeta} ^2
   + 6  \Lambda a ^3 {\tilde \zeta} ^3 + 3 \Lambda  a ^3 {\tilde \zeta} \,,\label{obviouspol}
\ee
where ${\tilde \zeta}$ stands for ${\tilde \zeta}(a)$. To study this cubic polynomial
inequality, it is convenient to introduce rescaled variables defined as
follows:
\begin{align} \label{eq:deffrak}
 \mathfrak{a} &\equiv  a \| \nab^{\rm L} \fett{v}^{\rm (init)} \| \,, \qquad
 \lambda \equiv   \Lambda  / \| \nab^{\rm L} \fett{v}^{\rm (init)}\|^3\,,
 \qquad    \zeta(\mathfrak{a})  \equiv  \tilde \zeta(a) \,.
\end{align}
The polynomial inequality \rf{obviouspol} becomes then
\be
 \label{pol}
     p_{\lambda}(\mathfrak{a}, \zeta) \equiv 6 \left( 1+ \lambda  \mathfrak{a}^3 \right)
    \zeta^3
+ 6  \left( 2+ \lambda  \mathfrak{a}^3 \right) \zeta^2
   +  \left( 3 \lambda  \mathfrak{a}^3 -1 \right) \zeta +  \mathfrak{a}  \geq 0 \,.
\ee
Sketches of the cubic polynomial $p_{\lambda}(\mathfrak{a}, \zeta)$ for $\lambda=0$ and for $\lambda \neq 0$ are respectively given in Figs.\,\ref{fig:analEdS} and~\ref{fig:analvaryomega}.

The cubic polynomial  $p_{\lambda}(\mathfrak{a}, \zeta)$
 has at least one real root. For
$\mathfrak{a}=0$ it has three roots: zero, a positive root and a negative
root. Since the displacement, and thus $\zeta$, 
have to vanish at time $\mathfrak{a}=0$,
the only physically relevant root of the cubic polynomial is zero. Now,
suppose we let the (rescaled scale-factor) time $\mathfrak{a}$ be small and
positive, then the physically relevant root moves
from zero to a small positive value
$\zeta_2 (\mathfrak{a})\approx \mathfrak{a}$. The polynomial
$p_{\lambda}(\zeta)$ is then strictly positive over the whole open interval
$\left] 0,\zeta_2 (\mathfrak{a}) \right[$, 
as required from the inequality~(\ref{pol}), 
and this is the only physically relevant branch of positivity.
Hence, within the interval $[0,\mathfrak{a}]$, the values of
the generating function $\zeta$ are bounded by $ \zeta_2 (\mathfrak{a})$.
As we increase the value of $\mathfrak{a}$, this boundedness property will hold
as long as there are two positive  roots in $\zeta$  for the 
polynomial $p_{\lambda}(\mathfrak{a},\zeta)$. This is true until the two 
positive roots merge into a double root and then turn 
into complex roots. The value of the rescaled
time for which \rf{pol} has a double root are the
 zeros of its discriminant, i.e.,
\begin{align} \label{eq:disc}
  \Delta({\mathfrak{a}},\lambda) &\equiv 12 \left(  14 -684 {\mathfrak{a}} -81 {\mathfrak{a}}^2 -76 {\mathfrak{a}}^3 \lambda -702 {\mathfrak{a}}^4 \lambda -162 {\mathfrak{a}}^5 \lambda
    +75 {\mathfrak{a}}^6 \lambda^2 
 - 81 {\mathfrak{a}}^8 \lambda^2 + 90 {\mathfrak{a}}^9 \lambda^3
    +90 {\mathfrak{a}}^{10} \lambda^3 -27 {\mathfrak{a}}^{12} \lambda^4  \right) =0 \,.
\end{align}
Because $\Delta(0,\lambda) >0$ and the highest-order term, proportional to
${\mathfrak{a}}^{12}$ has a negative coefficient, it immediately follows by
continuity that the discriminant equation has at least one positive root. 
For our purposes we need the \emph{smallest} of such positive roots, since it is an
upper bound for the physically relevant branch of $p_{\lambda}(\zeta) \geq 0$:
\be
\label{smallestroot}
T(\lambda) \equiv \inf_i \,\mathfrak{a}_i \,\,\big |  \,\,\Delta({\mathfrak{a}_i},\lambda)
=0\,,
\ee
called here  the critical time. (In Section~\ref{ss:obtainingtoflambda} we
shall show how $T(\lambda)$ can be calculated.)

The importance of the critical time lies in the following result: \textit{for any
$|\mathfrak{a}| < T(\lambda)$ the time-Taylor series for the gradient of the
displacement (now expressed in terms of the rescaled time) is convergent.}
(Here $|\mathfrak{a}|$ denotes the modulus of the rescaled time
$\mathfrak{a}$ which can now take also complex values.)
Indeed, it follows from~(\ref{eq:generate}) and~(\ref{eq:deffrak}) that
\be \label{eq:generatefrak}
  \zeta(\mathfrak{a}) = \sum_{s=1}^\infty \zeta ^{(s)} \mathfrak{a} ^{s} \,,
\ee
where  $\zeta ^{(s)} = \| \nabLT \fett{\xi}^{(s)} \| / \| \nabLT
\fett{v}^{(\rm init)} \|^{s}  \geq 0$. If $\zeta(T)$ is bounded by some
constant $C$, we have
\be
  \zeta(T) = \sum_{s=1}^\infty \zeta ^{(s)} T^s \leq C \,.
\ee
Since all the terms in the sum are non-negative, it follows that
\be
  \zeta ^{(s)} \leq C T^{-s} \,,
\ee
and thus
\be \label{eq:geometric}
  \zeta(\mathfrak{a}) \leq C \sum_{s=1}^\infty  \left( \frac{\mathfrak{a}}{T} \right)^s \,.
\ee
For $| \mathfrak{a} | < T$ the Taylor series~(\ref{eq:geometric}) is
bounded by a convergent geometric series.
We have thus shown the convergence of the
Taylor series in the complex time plane inside a disk, centered at the origin,
$\mathfrak{a} = 0$ of radius at least $T(\lambda)$. The actual radius of
convergence of the Taylor series, called the \textit{radius of analyticity}, for which $T(\lambda)$ is only a lower bound, in general, can only be determined numerically.

\section{Further results on analyticty}
\label{s:further}

\subsection{The lower bound on the radius of analyticity and its dependence on the
  cosmological constant}
\label{ss:obtainingtoflambda}

The  lowest positive root $T(\lambda)$ of the discriminant equation
\rf{eq:disc} gives us a lower bound
on the radius of analyticity of the time-Taylor series for the Lagrangian
map. This  discriminant equation  is of twelfth degree in the
rescaled time variable $\mathfrak{a}=  a \| \nab^{\rm L} \fett{v}^{\rm (init)}\|$ 
and we have not been able to solve it explicitly by radicals.
We can however solve the discriminant equation perturbatively
for small and large rescaled cosmological
constant $\lambda= \Lambda  / \| \nab^{\rm L} \fett{v}^{\rm (init)}\|^3$ and
numerically for other values.

For $\lambda =0$, we recover the results for an EdS
universe (Zheligovsky \& Frisch 2014). In that case the discriminant equation
reduces to  $\Delta({\mathfrak{a}},0) = 12 (  14 -684 {\mathfrak{a}} -81
{\mathfrak{a}}^2)=0$, whose lowest positive root is
\be \label{eq:TofLambda}
T^{\rm EdS} \equiv T(0)= 3 \sqrt{2} - {38}/{9} \simeq 0.0204 \,.
\ee
 For small
$\lambda$, it is then straightforward to obtain the $\lambda$-dependence of this root
\be
   T(\lambda) = T^{\rm EdS} - \left( T^{\rm EdS} \right)^3  \frac{52
    -333 T^{\rm EdS}}{342+81T^{\rm EdS}} \, \lambda + { O}(\lambda ^2)\simeq
   0.0204 -1.1197 \cdot 10^{-6} \lambda + { O}(\lambda ^2) \,.
\label{smalllambdaexp}
\ee

For large positive $\lambda$, we make the changes of variable
$\tilde{\mathfrak{a}} \equiv \mathfrak{a} \lambda ^{1/3}$  and
$\tilde{\Delta}(\tilde{\mathfrak{a}},\lambda) \equiv \Delta(\mathfrak{a},\lambda)/12$, 
and rewrite the discriminant equation as
\be
 \tilde{\Delta}(\tilde{\mathfrak{a}},\lambda) = P_4(\tilde{\mathfrak{a}}^3) +
 \lambda ^{-1/3}  P_{10}(\tilde{\mathfrak{a}}) + \lambda ^{-2/3}
 P_{8}(\tilde{\mathfrak{a}}) = 0 \,,
\ee
where
\be
P_4(\tilde{\mathfrak{a}}^3) =  14 -76 \tilde{\mathfrak{a}}^3 + 75
   \tilde{\mathfrak{a}}^6 + 90 \tilde{\mathfrak{a}}^9 -27
   \tilde{\mathfrak{a}}^{12}    \,,\quad P_{10}(\tilde{\mathfrak{a}}) = -684 \tilde{\mathfrak{a}} -702
    \tilde{\mathfrak{a}}^4 + 90 \tilde{\mathfrak{a}}^{10}     \,,\quad
    P_{8}(\tilde{\mathfrak{a}})  = -81 \tilde{\mathfrak{a}}^2 -162
\tilde{\mathfrak{a}}^5 - 81 \tilde{\mathfrak{a}}^8 \,.
\label{p4105}
\ee
The polynomial $P_4(\tilde{\mathfrak{a}}^3)$  has a positive double root at
$\tilde{\mathfrak{a}}_c  = 3^{-1/3}$, from which follows that, {to dominant order,} $T(\lambda)$ is proportional to $ 3^{-1/3} \lambda ^{-1/3}$. The
first subdominant correction is obtained by Taylor expanding the polynomial
$P_4$, $P_{10}$ and $P_8$ around $\tilde{\mathfrak{a}}_c$. This eventually
reveals the following large-$\lambda$ behaviour:
\be
T(\lambda) =  3^{-1/3} \lambda ^{-1/3} +{ O}(\lambda ^{-1/2}) \,.
\label{lasrgelambdaexp}
\ee

\begin{figure}
\includegraphics[width=\textwidth]{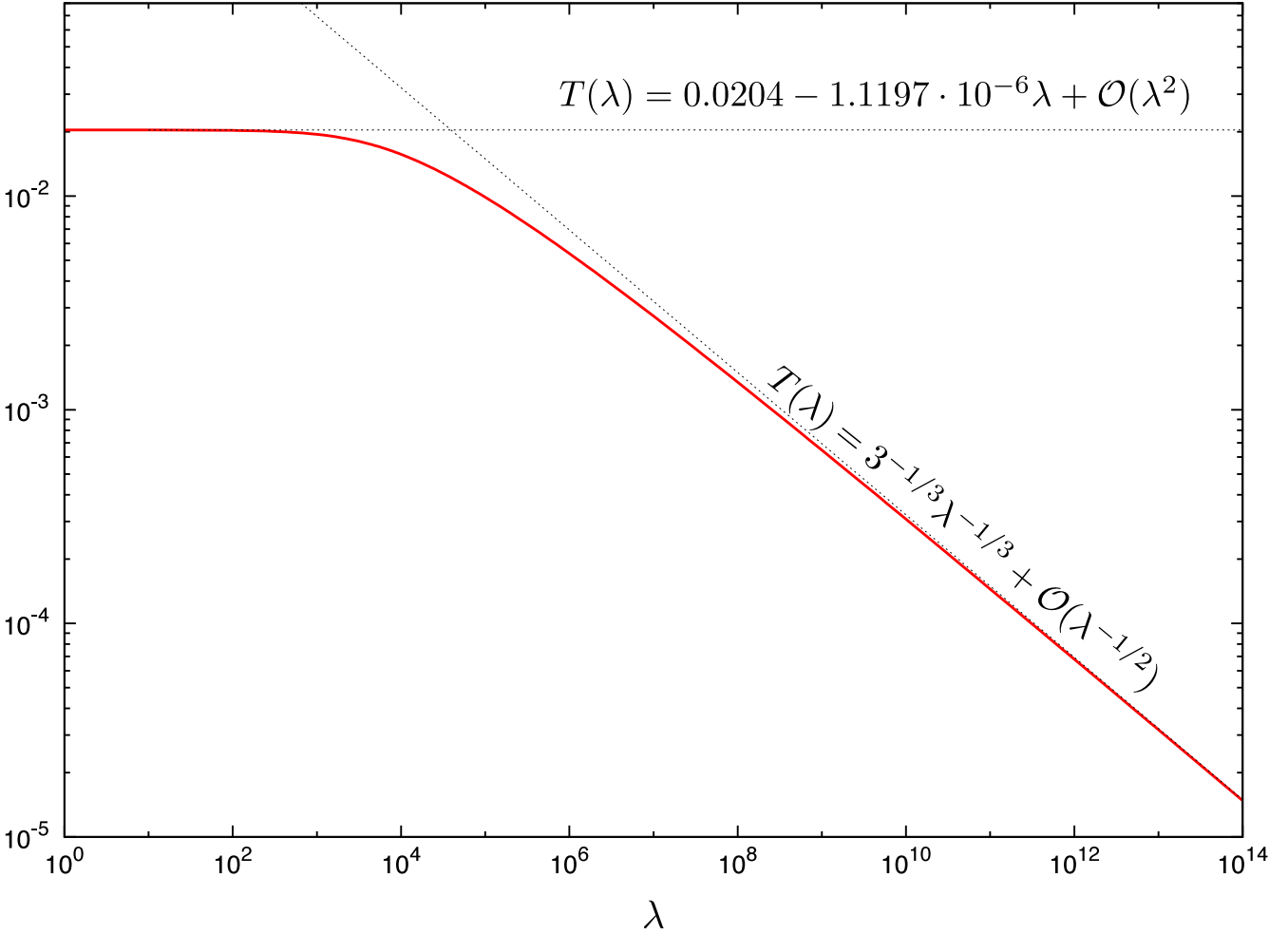}
\caption{Numerical solution of the discriminant equation \rf{eq:disc}.
The results agree with the small- and
large-$\lambda$ expansion \rf{smalllambdaexp} and~\rf{lasrgelambdaexp} in the
appropriate ranges.}
\label{fig:analTc}
\end{figure}

We have solved the discriminant equation \rf{eq:disc} numerically for more than one hundred values of
$\lambda$, suitably distributed between $10 ^{-2}$ and $10^{14}$. The results
are shown in Fig.~\ref{fig:analTc} and agree with the small- and
large-$\lambda$ expansion \rf{smalllambdaexp} and  \rf{lasrgelambdaexp} in the
appropriate ranges. Note that  $T(\lambda)$ is a monotonically decreasing
function of $\lambda$.

Finally, we observe that, within the functional framework of the space $\ell_1$ of
absolutely summable Fourier series, the precise values of the constants
appearing in the low-$\lambda$ expansion~\rf{smalllambdaexp} and the large-$\lambda$ expansion~\rf{lasrgelambdaexp}
can most likely be improved. This will  result in  longer time intervals of guaranteed
analyticity. For this one should adapt to the cosmological context the
detailed Fourier-space 
estimates found, for incompressible flow, in Section~2.3 of Zheligovsky \& Frisch (2014).

\subsection{Shell-crossing and analyticity}
\label{ss:shellcrossing}

In  section~\ref{sec:Lag}, we pointed out that the Lagrangian formulation and
the subsequent time-Taylor expansions as carried out in the present paper, are
invalid if the Jacobian $J$ vanishes during the interval of guaranteed
analyticity. The vanishing of the Jacobian corresponds to the crossing of
various fluid-particle trajectories. It is generally known as ``shell-crossing''.
We now show that during the time interval of guaranteed
analyticity, no shell-crossing can occur. Actually, denoting as usual the
Jacobian of the Lagrangian map by $J$, we shall show that
$|J-1|\le -37/9 + 3\sqrt 2 \approx 0.132$,
which trivially implies the non-vanishing of
the Jacobian (and not even a close call).

\begin{figure}
\includegraphics[width=\textwidth]{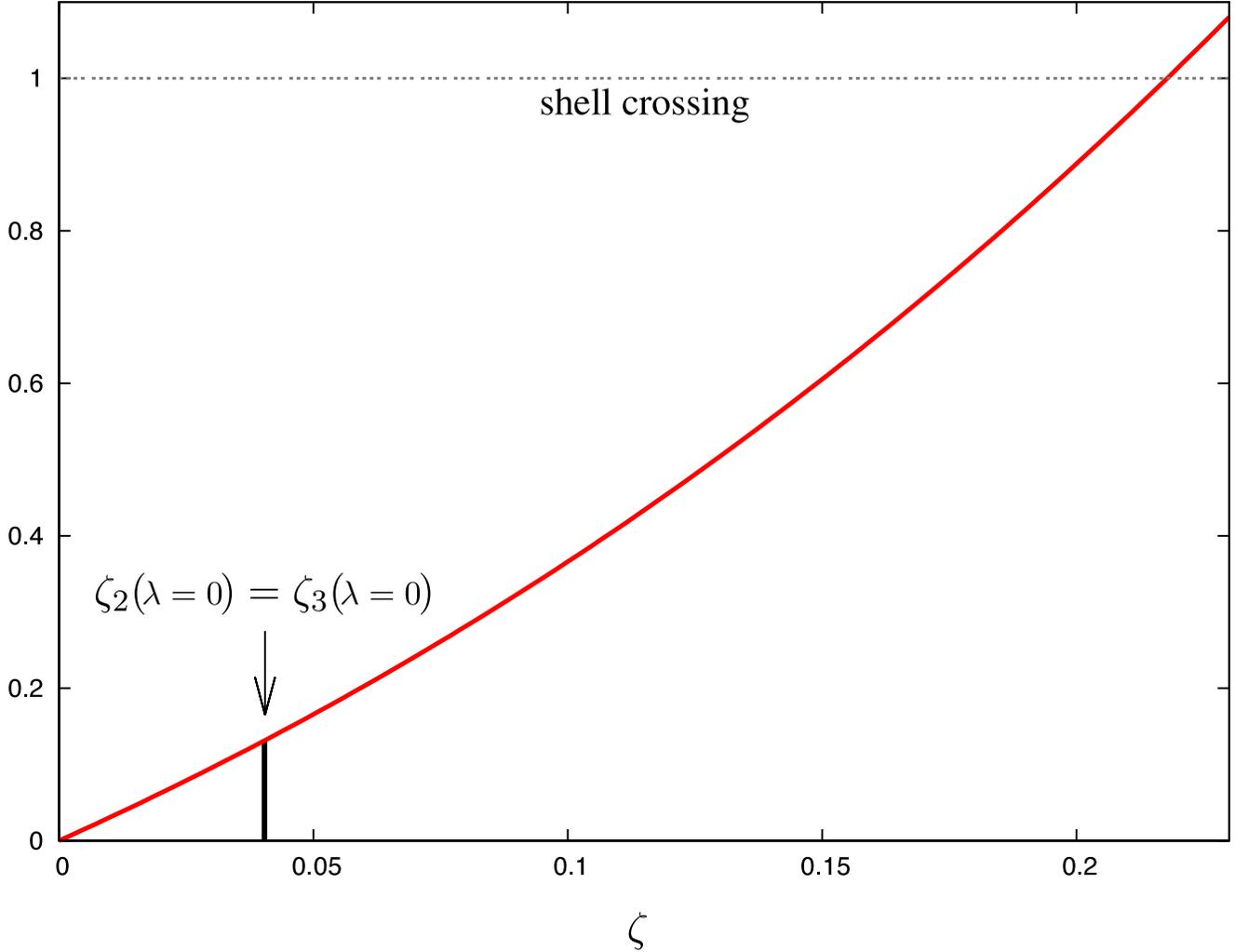}
\caption{Lower bound on the norm of $\left| J-1 \right| \le Q(\zeta) =6 \zeta^3 + 6 \zeta^2 + 3 \zeta$, see Eq.\,(\ref{boundjminus1}). The vertical thick line at roughly
$\zeta= 0.04044$ indicates the largest possible value of
the smallest positive root $\zeta_2(T(\text{\scriptsize $\lambda=0$}))$ of the polynomial $p_\lambda(\mathfrak{a},\zeta)$ for which analyticity is guaranteed. Since $\zeta_2(T(\lambda))$ is a decreasing function of $\lambda$ (see~(\ref{eq:zetaDoubleRoot})), it
follows that $|J-1| < 0.132$ for any value of $\Lambda>0$, which is well below the value unity where shell-crossing might occur (the dotted line).
}
\label{fig:jacobian}
\end{figure}

For this, we observe that the Jacobian can be written as follows
in terms of the displacement field
\be
J   = \det(I +\nabLT \fett{\xi})   = 1+   \nabLT \cdot
\fett{\xi} + \sum_{1 \leq i < j \leq 3}
\left[ (\nabL{i} \xi_i) \nabL{j} \xi_j - (\nabL{i} \xi_j
   ) \nabL{j} \xi_i \right]  + \det (\nabLT \fett{\xi}) \,,
\label{jdecomposition}
\ee
where $I$ denotes the identity matrix. Using the Taylor expansion
\rf{ansatzDispl} of the displacement in powers
of the $a$-time and the generating function $\zeta = \sum_{s=1}^\infty |
\nab^{\rm L} \fett{\xi}^{(s)} | a^s$, we obtain the following bound
\be
   \left| J-1 \right| \le 6 \zeta^3 + 6 \zeta^2 + 3 \zeta  \equiv Q (\zeta) \,.
\label{boundjminus1}
\ee
For convenience we have plotted the cubic polynomial $Q(\zeta)$
in Fig.~\ref{fig:jacobian}.

As we have seen in Section~\ref{sec:anal}, for any given rescaled cosmological constant $\lambda$ and rescaled time
$\mathfrak{a}$, the permitted values of the generating function $\zeta$ are
between zero and the smallest positive root $\zeta_2(\mathfrak{a})$ of the cubic polynomial
$p_\lambda(\zeta, \mathfrak{a})$, given by
\rf{pol}. We claim  that this root is an increasing function of
$\mathfrak{a}$ for fixed $\lambda$.
 Indeed,
$p_\lambda(\mathfrak{a},\zeta_2(\mathfrak{a})) =0 $,  over the whole
range of relevant  $\mathfrak{a}$ values. By differentiating it with respect to
$\mathfrak{a}$, we obtain:
\be \label{eq:dpda}
  \frac{\partial p_\lambda(\mathfrak{a},\zeta_2(\mathfrak{a}))}{\partial \mathfrak{a}} = \frac{\partial
    \zeta_2(\mathfrak{a})}{\partial \mathfrak{a}} \frac{\partial p_{\lambda}(\mathfrak{a},\zeta)}{\partial
    \zeta} \Big. \Big|_{\zeta= \zeta_2(\mathfrak{a})} + 1
  + 3 \lambda \mathfrak{a}^2 \left( 6 \zeta_2^3 + 6 \zeta_2^2 + 3\zeta_2 \right) = 0 \,.
\ee
Since  $[1 + 3 \lambda \mathfrak{a}^2(6 \zeta_2^3 + 6 \zeta_2^2 + 3
\zeta_2)]>0$ for $\zeta_2 \ge0$, we infer that
\be
\frac{\partial
    \zeta_2(\mathfrak{a})}{\partial \mathfrak{a}} \frac{\partial p_{\lambda}(\mathfrak{a},\zeta)}{\partial
    \zeta} \Big. \Big|_{\zeta= \zeta_2(\mathfrak{a})} <0\,.
\label{nearlythere}
\ee
The positivity of $\frac{\partial \zeta_2(\mathfrak{a})}{\partial
  \mathfrak{a}}$
follows from $\frac{\partial p_{\lambda}(\mathfrak{a},\zeta)}{\partial
    \zeta} \Big. \Big|_{\zeta= \zeta_2(\mathfrak{a})} <0$. The latter
is a consequence of the fact that the polynomial
$p_{\lambda}(\mathfrak{a},\zeta)$ has three real roots, for fixed $\lambda$ and  fixed
$\mathfrak{a}$ in the range of analyticity. Inspection of \rf{pol}
shows that their product is negative, that one ($\zeta_1$) is negative
and that two ($\zeta_2$ and $\zeta_3$ ) are
positive. The derivative $\frac{\partial p_{\lambda}(\mathfrak{a},\zeta)}{\partial
    \zeta}$ is obviously positive for large $\zeta >0$ and changes
sign somewhere between $\zeta_2$ and $\zeta_3$. Since
$\zeta_2<\zeta_3$, this derivative is negative at $\zeta_2$. This
proves the monotonic increase of $\zeta_2$ with $\mathfrak{a}$.

Next, we use \rf{boundjminus1} to bound $|J-1|$ by  $6 \zeta_2^3 + 6 \zeta_2^2 + 3 \zeta_2$. The largest possible
value of $\zeta_2$ is obtained at the critical value $\mathfrak{a}=T(\lambda)$
where the polynomial $p_\lambda(\mathfrak{a},\zeta)$ has
a double root in $\zeta$. This value can be obtained by demanding the
simultaneous vanishing of  $p_\lambda(\mathfrak{a},\zeta)$ and of its
derivative with respect to $\zeta$, i.e.,
$p_\lambda(\zeta_2)=p_{\lambda}'(\zeta_2) =0$ and $\zeta_2 \geq 0$,
where a prime denotes a partial derivative with respect to $\zeta$.
From these conditions we obtain
\be \label{eq:zetaDoubleRoot}
  \zeta_2(T(\lambda)) = \zeta_3(T(\lambda)) = \frac{-4 -2 \lambda T^3
   + \sqrt{2} \sqrt{9+ \lambda T^3 \left( 6 -\lambda T^3 \right)}}{6 + 6 \lambda T^3} \,.
\ee
Since this is a decreasing function of $\lambda$, it is
bounded by its EdS value ($\lambda =0$),
where the latter is $\zeta_2= - \frac 2 3 +
\frac{1}{6} \sqrt{18 } \simeq 0.04044$. It follows that $|J-1| <
0.132$, which is well below the value unity where shell-crossing might
occur (see the vertical thick line in Fig.~\ref{fig:jacobian}).

\subsection{Analyticity in the linear growth function $D$}
\label{ss:D}

As already stated in Section~\ref{sec:EPS}, it could be of interest  in
cosmological studies to use as time variable, not the cosmic scale factor $a$, but
the linear growth function $D(a)$, which is the growing solution of
\rf{eq:linEuler}. With the normalisation 
 $\lim_{a\to 0} D(a)/a
=1$, the linear growth function is given by
\be
D(a) = a \sqrt{1+ \Lambda a^3} \, {}_2F_1\!\left( \frac 3 2, \frac 5 6, \frac{11}{6}, -
\Lambda a^3 \right) \,,
\ee
where ${}_2F_1$ is the Gauss hypergeometric
function. This solution is analytic around
$a\!=\!0$ and  has the small-$a$ expansion  $D(a) = a - (2/11)\,\Lambda a^4 +
{ O}(a ^7)$. (The full expression can be obtained from the Taylor
expansion of the Gauss hypergeometric
function, cf.~Gradshteyn \& Ryzhik (1965),  pp.~1039--1045.)
As $D(a)$ is analytic around $a=0$ and also its derivative
$\partial D /\partial a$ does not
vanish at $a=0$,
 it follows that $D(a)$ is invertible at around $a=0$.
The inverse linear growth function $a(D)$ is also analytic and has the following
low-order expansion: $a(D) = D + (2/11)\,\Lambda D^4 +{ O}(D ^7)$.

We have shown in Section~\ref{sec:anal} that the Lagrangian map
$\fett{x}(\fett{q},a)$ is an analytic function of $a$, near $a=0$. The
composition $\fett{x}(\fett{q},a(D))$ of the Lagrangian map with the inverse
linear growth function is  also analytic in $D$. (This follows basically from the
observation that analytic functions are complex-differentiable functions
and the use of the chain rule.)
We remark that this argument is not valid when the cosmic time is chosen 
as time variable, instead of $D$, as the relation between $a$ and $t$ 
is not analytic near the origin.

Can we obtain simple recursion relations for the $D$-Taylor coefficients of
the displacement $\fett{\xi}$, as we have done in Section~\ref{sec:rec}?
Recursion relations, yes, but simple, no. Indeed, from
\rf{eq:main2}--\rf{eq:main}, we can derive a set of Lagrangian Euler--Poisson
equations in the time variable $D$. However the $D$-dependence is then
essentially not polynomial, but will involve the full inverse linear growth
function $a(D)$, whose Taylor expansion has an infinite number of terms. As a
consequence, the recursion relations will be very involved. If one really
wants to reexpress the displacement as a power series in $D$, it is much
simpler to just substitute the Taylor series for $a(D)$, obtained by inversion
of the Taylor series of $D(a)$ into the expansion \rf{ansatzDispl} in powers
of $a$. Note that, with the latter strategy, there is no need to solve a
succession of time differential equations for the $n$th order growth function
(usually denoted with $E$, $F$, \dots), because the actual expansion parameter
is precisely the cosmic scale factor, used as a time variable.

\subsection{Effect on analyticity of the inclusion in the Friedmann
  equation of curvature and radiation terms}
\label{ss:CR}

Inflation theory and observational evidence since the late nineties
favour a flat Universe with zero curvature [see, e.g.,  Linde~1984 and
Netterfield et al.\ (Boomerang Collaboration) 2002].
It is nevertheless
of interest to point out that our analyticity result for a $\Lambda$CDM
Universe survives when a small amount of curvature is introduced  in
the  Friedmann equation \rf{eq:Friedmann}. The latter then becomes
\be
 \left( \frac{\dot{a}}{a}\right)^2 =  a^{-3} + k a ^{-2} +\Lambda \,,
\label{friedmanncurvature}
\ee
where $k$ is here not limited to being $0,\, \pm 1$ (although according
to Mukhanov~(2005),
the only trivial background geometries which do not spoil exact 
homogeneity and isotropy are indeed the ones with curvature $0,\, \pm 1$).
We have repeated the analysis of
Sections~\ref{sec:EPS}--\ref{sec:anal}, with this modified Friedmann
equation and found that very little is changed. The slaving conditions
\rf{slaving} are unchanged. The recursion relations \rf{eq:nabRec} are
minimally modified.
 The polynomial inequality \rf{pol} in the generating
function
$\zeta$ takes now the following form in rescaled variables:
\be
   P_{\lambda, \mathfrak{K}}(\mathfrak{a},\zeta) \equiv 6 \left( 1+ \lambda {\mathfrak{a}}^3  + 2
   \mathfrak{K} {\mathfrak{a}} \right) \zeta^3 + 6  \left( 2 + \lambda {\mathfrak{a}}^3  + \frac{8}{33}
   \mathfrak{K} {\mathfrak{a}} \right) \zeta^2
   +  \left( 3 \lambda {\mathfrak{a}}^3 + \frac{12}{7} \mathfrak{K} {\mathfrak{a}} -1 \right) \zeta + {\mathfrak{a}} \geq 0 \,,
\label{polk}
\ee where $\mathfrak{K} =|k|/\| \nab^{\rm L} \fett{v}^{\rm (init)} \|$, and
$\mathfrak{a} = a \| \nab^{\rm L} \fett{v}^{\rm (init)} \|$ and $\lambda =
\Lambda / \| \nab^{\rm L} \fett{v}^{\rm (init)}\|^3$ as before.  Analyticity in the
(rescaled) cosmic scale factor $\mathfrak{a}$ is proved as before, by studying the
discriminant equation associated to the cubic polynomial 
$P_{\lambda,\mathfrak{K}}(\mathfrak{a},\zeta)$. Indeed, this is again an equation of
twelfth degree in $\mathfrak{a}$, whose lowest and highest degree terms,
unchanged from the $\Lambda$CDM case, are respectively positive and negative,
thus implying the existence of at least one positive root.

All this is not surprising: Actually, the curvature term in the Friedmann
equation involves the inverse second power of the cosmic scale factor, which
for $a \to 0$ is subdominant with respect to the matter term $a ^{-3}$ and can
thus not bring about much change. Very
different would be the inclusion in the background  of \textit{radiation effects}, stemming from a
black-body distribution of photons. This adds a  term in
the Friedmann equation, proportional to the inverse fourth power of the scale
factor, which drastically changes everything: slaving is lost, analyticity around
$a=0$ is lost, etc. The current favoured $\Lambda$CDM model has a background radiation
density well below the matter density at decoupling and can
actually be discarded after decoupling. When relating
our results to numerical results or observations, it may be more convenient
to use  the cosmic time $t$ rather than the $a$-time. Of course, when performing that
change of variable, the radiation term in the Friedmann equation does not need to be omitted.

\section{Concluding remarks}\label{sec:conc}

In a Eulerian 
 framework space and time are intertwined by Galilean
transformations (in the Newtonian approximation). It follows that, with the
limited spatial smoothness assumed in the present work -- roughly, that the initial
density fluctuations are slighly better than continuous in the spatial
variables -- one cannot obtain better than a corresponding limited temporal
smoothness. In a Lagrangian framework, as we have seen, the situation is
drastically different: provided we use an appropriate time variable, such as
the cosmic scale factor $a$ or the linear growth function $D$. Obtaining analyticity
in time of the Lagrangian trajectories requires then only a limited initial spatial
smoothness.

Even more surprising is that this analyticity is around the point $a=0$, which
seems to extend this utterly smooth birth of our structured Universe to the
very origin of time, well before decoupling of matter from photons. Of course,
this is just a property of the mathematical model used here to describe
structure formation, namely the Euler--Poisson equations in a $\Lambda$CDM
Universe. The interesting feature is that this model allows us to have a
well-posed problem, devoid of catastrophic behaviour near decoupling, 
provided we use the \textit{slaving conditions}~\rf{slaving} at $a=0$. 
These slaving conditions resemble those used to start $N$-body
simulations at early times, (i) small density fluctuations, (ii) an initial
curl-free peculiar velocity field, 
 related to the gravitational potential as
in the Zel'dovich approximation. That these slaving conditions can be extended
all the way to $a=0$ means, from the point of view of boundary layer analysis
(Cole~1968), 
that the matter-dominated era constitutes an outer solution
for which, to leading order, the boundary matching to the inner solution (describing
the primordial era) can be replaced by just an initial condition. This is true
as long as, in the Friedmann equation for the background evolution in the
matter-dominated era, we include only a matter term, a cosmological term and a
possible curvature term, but no radiation term.

We remind the reader that our way of proving analyticity in the
$a$-time is fully constructive. It rests indeed on a set of
novel all-order recursion relations \rf{eq:nabRec} for the time-Taylor
coefficients of the displacement of (fluid) particles. Not only are
these relations fully explicit, but they have a very specific
mathematical structure, allowing us to obtain bounds for all the
Taylor-coefficients and thus to establish analyticity by elementary means.

We now briefly indicate some of the possible future developments exploiting
the analyticity result and the recursion relations. We shall mention
two interlinked topics. One is about numerical integration of the Euler--Poisson
equations by a multi-step technique and the other one about
cosmological reconstruction.

The Taylor method as described in Sections~\ref{sec:rec} and
\ref{sec:anal} allows us to determine the solution of the
Euler--Poisson equations for any time $0\!<\!a\!<\!R(0)$, where $R(0)$ is the
radius of convergence of the Taylor series around $a\!=\!0$. (Here, it is
best not to work with the rescaled \mbox{time $\mathfrak{a}$.)} A
procedure, inspired from the Weierstrass analytic continuation
technique for analytic functions, 
 may allow us to obtain the
solution beyond the time $R(0)$. For this, we select a sequence of
times $0<a_1<a_2<a_3<\ldots$ in such a way that
$a_{n+1}$ is within the disk of convergence of the time-Taylor series
around the time $a_n$, that is such that $|a_{n+1}- a_n|< R(a_n)$,
where $R(a)$ is the radius of convergence of the Taylor expansion of
the solution around time $a$. What we here call the solution is not
just the Lagrangian map, but includes also the density which is
controlled by the inverse of the Jacobian. Indeed, if the Jacobian
vanishes, shell-crossing takes place and the Lagrangian map ceases to
be invertible. Invertibility is essential, because at the end of each
step, we must revert to a Eulerian description to be able to make a
fresh start.  Methods combining a Lagrangian step with a reversion to
Eulerian coordinates are called semi-Lagrangian. A detailed
description of a semi-Lagrangian method, called the Cauchy--Lagrangian
method, which makes use of Cauchy's Lagrangian formulation of the
ideal fluid dynamics can be found in Podvigina \textit{et al.}~(2015). 

When dealing with the cosmological Euler--Poisson equations, the
method of Podvigina \textit{et al.}~(2015) must be suitably adapted. One
of the new difficulties stem from the non-autonomous character of our
Euler--Poisson equations: the time appears explicitly in the equation.
As a consequence, the Euler--Poisson equations are not invariant under $a$-time
translations, but this problem can be circumvented.

In so far as the regime described by the Euler--Poisson equations does
not include multi-streaming, a Cauchy--Lagrangian numerical method will
not be able to address the same questions as $N$-body simulation
techniques.  There is however one important problem for which it
appears well suited, namely the reconstruction of the dynamical
history of the Universe from present-epoch galactic surveys (Frisch {\it et al.}~2002).
It has been shown by Brenier \textit{et al.}~(2003) 
that the solution to the Euler--Poisson equations is uniquely determined by two
boundary conditions, the slaving conditions \rf{slaving} at the
initial time and the density of matter at the current epoch. The
latter is obtained from large galactic surveys.  Euler--Poisson
reconstruction is an extension of the variational $N$-body
reconstruction, introduced by Peebles~(1989) 
for handling galaxy data on relatively small scales, such as that of the Local
Group. Euler--Poisson reconstruction, which is meant for significantly
larger scales where a continuum description is applicable and where
multi-streaming can be ignored, does not suffer from the
non-uniqueness problem of the variational $N$-body reconstruction.  So
far, Euler--Poisson reconstruction has not used the full solutions of
the Euler--Poisson equations but just low-order Lagrangian perturbation
approximations, such as the Zel'dovich approximation or the next
order, denoted by 2LPT. Within such approximate frameworks,
reconstruction becomes a Monge optimal transport problem
(Frisch {\it et al.}~2002) with quadratic cost, which, after discretisation,
can be solved by very efficient assignment algorithms such as the
auction method of Bertsekas~(1992). 
Alternatively, one can use Brenier's theorem (Brenier~1987) and solve an
equivalent  3-D Monge--Amp\`ere
equation non-linear PDE by iterative techniques  (Zheligovsky, Podvigina \& Frisch 2010). 
Reconstruction by such methods is known as 
Monge--Amp\`ere--Kantorovich (MAK).  We propose to use the MAK reconstruction
as a starting point of full Euler--Poisson reconstruction, applying an
iterative Newton-type method in which at each stage a Euler--Poisson
initial value problem is solved by a Cauchy--Lagrangian method.

Finally, we remind the reader that our proofs were given entirely within a
Newtonian framework. In that framework the time-analyticity results revealed
a strong asymmetry between space and time in Lagrangian coordinates.
The reason for that strong asymmetry is the ``missing'' convective term in the
Newtonian fluid equations, when formulated in Lagrangian coordinates.
(In the notation of our Eulerian approach in Section~\ref{sec:EPS},
the convective term is $\fett{v}\cdot \!\nabx\fett{v}$.)
An interesting question is whether our time-analyticity results could be
generalised to a Lagrangian coordinates formulation of a curl-free dust-fluid
in General Relativity.
It is well-known that the so-called synchronous-comoving coordinate system
corresponds to a relativistic Lagrangian frame of reference, and,
similar to the Lagrangian coordinates approach in the Newtonian framework,
there is indeed no convective term in the relativistic equations of motion
(in the so-called ADM approach; see e.g.~Matarrese \& Terranova~1996).
These relativistic fluid equations are however decorated with Ricci-tensors that
involve (single and double) spatial gradients acting on the metric
(and on its inverse),
and this could well imply a limited temporal smoothness for
limited spatial smoothness.
{Besides that unsolved problem, to obtain a time-analyticity
result in General Relativity along lines similar to those of the present paper, 
it seems necessary to first obtain explicit all-order recursion relations.}
A possible starting point for such investigations could be
the Lagrangian equations recently obtained by Alles {\it et al.}~(2015).

\section*{Acknowledgments}

We thank T.~Buchert, A.~Sobolevski\u{\i}, A.~Wiegand and V.~Zheligovsky  for useful
discussions and/or comments on the manuscript. 
U.F.~and B.V.~acknowledge the support of the
F\'ed\'eration Wolfgang D\"oblin (CNRS, Nice).
C.R.~acknowledges the support of the individual fellowship RA 2523/1-1 
from the German research organisation (DFG).


\appendix

\section{Derivation of the peculiar Euler--Poisson equations with cosmological constant}
\label{b:Nedsequations}

Here, although entirely elementary, we highlight the derivation of
the peculiar Euler--Poisson equations~(\ref{eq:Euler}--\ref{eq:Pois}) for
the sake of completeness.
We begin with the (full) Euler--Poisson equations~(\ref{eq:Eulermn}--\ref{eq:lambdapoisson}), which we repeat here for convenience:
\begin{align}
& \label{app:Eulermn} \partial_t \tilde{\fett{U}} + \left( \tilde{\fett{U}} \cdot \nabr \right) \tilde{\fett{U}} =  - \nabr  \tilde{\phi}_g \,, \\
& \partial_t \tilde\varrho +\nabr \cdot \left( \tilde\varrho \tilde{\fett{U}}\right) = 0 \,, \\
\label{app:poisson}
& \nabr^2  \tilde{\phi}_g = 4\pi G \tilde\varrho - 3\Lambda \,.
\end{align}
Here $\tilde{\fett{U}}= \tilde{\fett{U}} (\fett{r},t)$ denotes the velocity field,
 $\tilde{\phi}_g = \tilde{\phi}_g (\fett{r},t)$ denotes the cosmological potential,
$\tilde\varrho=\tilde\varrho (\fett{r},t)$ is the fluid density, and here the cosmological constant is $3\Lambda$.
Let us solve these equations for the {\em background variables} in an exactly homogeneous and isotropic Universe, where $\tilde\varrho=\tilde{\bar\varrho} (t)$,
$\tilde{\fett{U}}=\tilde{H}(t) \,\fett{r} \equiv ({\dot{ a}}/{ a})\fett{r}$, and $\tilde{\phi}_g =
\tilde{\bar{\phi}}_g$. We substitute these expressions into the above equations.
For the divergence of Eq.\,(\ref{app:Eulermn}), we then have
\be \label{eq:nablabackEulermn}
  3 \dot{\tilde{H}}  + 3 \tilde{H}^2 = - \nabr^2 \tilde{\bar \phi}_g \,,
\ee
whereas for Eq.\,(\ref{app:poisson}), we have at the background level
\be \label{backgroundPoisson}
\nabr^2  \tilde{\bar \phi}_g = 4\pi G \tilde{\bar \varrho} - 3\Lambda \,.
\ee
By substituting the last expression into Eq.\,(\ref{eq:nablabackEulermn}), we obtain
\be \label{secondFriedmann}
\frac{\ddot a} {a} = - \frac {4 \pi G} {3} \tilde{\bar \rho} + \Lambda \,.
\ee
 This is the usual form of the second Friedmann equation for a pressureless fluid.
In the units after Eq.(\ref{eq:Friedmann}) in the main text, the above is $\ddot a/a = -a^{-3}/2+ \Lambda$, whose solution is
   $a(t) = \Lambda^{-1/3} [\sinh (3\sqrt{\Lambda}t/2)]^{2/3}$.

Now we take the fluctuations of our variables into account, i.e., we demand
\be \label{app:decomp}
  \tilde\varrho = 
   \tilde{\bar\varrho}(t) \left[ 1 + \tilde\delta \right] \,, \qquad
  \tilde{\fett{U}} = \frac{\dot{a}(t)}{a(t)} \fett{r} + a(t)\tilde{\fett{u}} \,, \qquad
   \tilde{\phi}_g  = \tilde{\bar\phi}_g + \tilde{\varphi}_g \,,
\ee
In order to write the Poisson equation for the {\em peculiar} potential $\tilde{\varphi}_g$,
 we use Eqs.\,(\ref{app:poisson}) and~(\ref{backgroundPoisson}), and obtain
\be \label{pecpot}
\nabr^2 \tilde{\varphi}_g = \nabr^2  \tilde{\phi}_g  - \nabr^2  \tilde{\bar \phi}_g =  4\pi G \left( \tilde\rho - \tilde{\bar \rho} \right)
  =  4\pi G \tilde{\bar \varrho} \tilde\delta  \,.
\ee
Observe that $\Lambda$ has dropped out in the peculiar Poisson equation, i.e., $\Lambda$ does only affect the background evolution of the Poisson equation.

Let us now derive the peculiar evolution equation of Eq.\,(\ref{app:Eulermn}) by use
of~(\ref{app:decomp}). We have
\be \label{eq1}
\partial_t \Big( \frac {\dot a}{a}\fett r + a\tilde{\fett u} \Big) + \left[ \left( \frac {\dot a}{a}\fett r + a\tilde{\fett u} \right) \cdot \nabr \right]    \left( \frac {\dot a}{a}\fett r + a \tilde{\fett u} \right) = -\nabr \left(\tilde{\bar \phi}_g +  \tilde{\varphi}_g\right) \,.
\ee
Let us do the calculations in (\ref{eq1}) step by step. The
l.h.s.~of~(\ref{eq1}) can be written as
\begin{align} \label{eq2}
\text{l.h.s.} &=  \frac {\ddot a}{a}\fett r - \frac {{\dot a}^2}{a^2} \fett r 
+ \dot a \tilde{\fett u} + a \partial_t \tilde{\fett u} 
+ \frac {{\dot a}^2}{a^2}  \big(\fett r \cdot \nabr \big) \fett r 
+ \dot a \big(\tilde{\fett u} \cdot \nabr \big) \fett r 
+ \dot a  \big(\fett r \cdot \nabr \big) \tilde{\fett u} 
+ a^2 \big(\tilde{\fett u} \cdot \nabr  \big) \tilde{\fett u}   \nonumber \\
 &= \frac {\ddot a}{a}\fett r    - \frac {{\dot a}^2}{a^2} \fett r 
+ \dot a \tilde{\fett u} +  a \partial_t \tilde{\fett u} 
+ \frac {{\dot a}^2}{a^2} \fett r + \dot a \tilde{\fett u}  
+ \dot a \big( \fett r \cdot \nabr\big) \tilde{\fett u}  
+ a^2 \big(\tilde{\fett u} \cdot \nabr \big) \tilde{\fett u} \,.
\end{align}
We thus obtain
\be
\label{eq3}
a \partial_t \tilde{\fett u} + a^2 \big(\tilde{\fett u} \cdot \nabr \big) \tilde{\fett u} + 2 \dot a \tilde{\fett u}  + \dot a \big(\fett r \cdot \nabr \big) \tilde{\fett u}  =
  - \nabr \tilde{\varphi}_g.
\ee
Now, we change the dependence of the above functions from $\fett{r}$ to $\fett{x} = \fett{r}/a$. For an arbitrary function $\tilde f(t,\fett r) = \tilde f (t, a(t) \fett x)$, we have
\be \label{eq4}
\partial_t \tilde f |_{\fett x\, \text{fixed}} = \partial_t \tilde f |_{\fett r\, \text{fixed}} + \dot a \big(\fett x \cdot \nabr\big) \tilde f |_{t\, \text{fixed}} \,.
\ee
Doing the same for our Eqs.\,(\ref{eq3}), and note that $\nabr = \frac {1}{a} \fett{\nabla}_{\fett{x}}$, we obtain
\be \label{eq6}
a \partial_t \tilde{\fett u} |_{\fett r\, \text{fixed}} - \dot a \big(\fett r \cdot \nabr \big) \tilde{\fett u} + a^2 \big( \tilde{\fett u} \cdot  \frac {1}{a} \fett{\nabla}_{\fett{x}}\big) \tilde{\fett u}  + 2 \dot a \tilde{\fett u} + \dot a \big(\fett r \cdot \nabr\big) \tilde{\fett u} = -   \frac {1}{a} \fett{\nabla}_{\fett{x}}  \tilde{\varphi}_g.
\ee
Then, we divide by $a$\, and obtain
\be
\partial_t \tilde{\fett u} + \big(\tilde{\fett u} \cdot \fett{\nabla}_{\fett{x}} \big) \tilde{\fett u} + 2 \frac {\dot a}{a} \tilde{\fett u} = - \frac {1}{a^2}\fett{\nabla}_{\fett{x}} \tilde{\varphi}_g \,.
\ee
The above equation, for an EdS universe was derived by Brenier {\it et al.}~(2003).
Note that we have corrected a typo in their equation~(A12).

\label{lastpage}


\begin{thebibliography}{}


\bibitem[Alles {\it et al.} 2015]{Alles:2015vua}
  Alles A., Buchert T., Roumi F.A., Wiegand A., 2015, Phys.\ Rev.\ D, 92, 023512


\bibitem[Arfken 2005]{arfken}
Arfken G.B., Weber H.J., 2005, Mathematical Methods for Physicists,
Elsevier, Amsterdam



\bibitem[Bernardeau 2013]{Bern13}
Bernardeau F., 2015, in Deffayet C., Peter P., Wandelt B., Zaldarriaga M., Cugliandolo L., eds, Lecture Notes of the Les Houches Summer School 2013,
Post-Planck Cosmology, Vol. 100, Oxford Univ. Press, p.\ 13

\bibitem[Bernardeau {\it et al.}~2002]{Bernardeau:2001qr}
Bernardeau F., Colombi S., Gaztanaga E., Scoccimarro R., 2002, Phys. Rep.,
367,1

\bibitem[Bertsekas 1992]{Bertsekas1992}
Bertsekas D.P., 1992,
Comput. Optim. Appl., 1, 7

\bibitem[Bildhauer]{Bildhauer}
Bildhauer S., Buchert T., Kasai M., 1992, A\&A, 263, 23


\bibitem[Bouchet {\it et al.}~1992]{Bo92}
Bouchet F.R., Juszkiewicz R., Colombi S., Pellat R., 1992, ApJ, L5

\bibitem[Bouchet {\it et al.}~(1995)]{Bouchet95}
 Bouchet F.R., Colombi S., Hivon E.,  Juszkiewicz R., 1995, A\&A, 296, 575

\bibitem[Brenier~1987]{Brenier1987}
Brenier Y., 1987, C.R. Acad. Sci. Paris I, 305, 805


\bibitem[Brenier {\it et al.}~2003]{Brenier}
 Brenier Y., Frisch U.,  H{\'e}non M., Loeper G., Matarrese S., Mohayaee R., Sobolevski A., 2003, MNRAS, 346, 501

\bibitem[Buchert 1989]{Buchert:1989xx}
  Buchert T., 1989,
  A\&A, 223, 9

\bibitem[Buchert 1992]{Buchert92}
Buchert T., 1992, MNRAS, 254, 729

\bibitem[Buchert~1995]{Buchert:1995bq}
Buchert T., 1995, in Bonometto S., Primack J. R., Provenzale A., eds, Proc.\ International School of Physics Enrico Fermi, Course Cxxxii, Dark Matter in the
Univ., Vol. 132, IOS Press Amsterdam, p.\ 543

\bibitem[Buchert \& Goetz 1987]{Buchert:1987xy}
  Buchert T., Goetz G., 1987,
  J.\ Math.\ Phys., 28, 2714


\bibitem[Buchert, Nayet \& Wiegand 2013]{Buchert:2013qma}
  Buchert T., Nayet C., Wiegand A., 2013, Phys.\ Rev.\ D, 87, 12, 123503

\bibitem[Buchert \& Ostermann 2012]{Buchert:2012mb}
  Buchert T., Ostermann M., 2012,
  Phys.\ Rev.\ D, 86, 023520

\bibitem[Catelan (1995)]{Catelan}
Catelan P., 1995, MNRAS, 276, 115


\bibitem[Cole~1968]{Cole1968}
  Cole J.D., 1968, Perturbation Methods in Applied Mathematics, Ginn/Blaisdell, Waltham

\bibitem[Crocce {\it et al.}~2006]{Crocce:2006ve}
  Crocce M., Pueblas S., Scoccimarro R., 2006,
  MNRAS, 373, 369

\bibitem[Demianski {\it et al.}~2005]{Demianski:2005is}
  Demianski M., Golda Z.A., Woszczyna A., 2005,
  Gen.\ Rel.\ Grav.,  37, 2063

\bibitem[Ehlers \& Buchert (1997)]{EhlersBuchert97}
Ehlers J., Buchert T., 1997, Gen. Rel. Grav., 29, 733

\bibitem[Enqvist \& Rigopoulos 2010]{Enqvist:2010ex}
  Enqvist K., Rigopoulos G., 2010,  J. Cosmol. Astropart. Phys., 1212, 005


\bibitem[Frisch \& Villone 2014]{FriVil}
Frisch U., Villone B., 2014,  Eur. Phys. J. H, 39, 325


\bibitem[Frisch {\it et al.}~2002]{Frisch:2001vw}
  Frisch U., Matarrese S., Mohayaee R., Sobolevski A., 2002,
  Nature,  417, 260


\bibitem[Gradshteyn \& Ryzhik 1965]{Gradshteyn}
Gradshteyn I. S., Ryzhik I. M., 1965, in Jeffrey A., ed., Table of Integrals Series and Products, 4th edn Academic Press, New York, p.\ 1039


\bibitem[Hamber \& Toriumi 2010]{Hamber:2010an}
  Hamber H.W., Toriumi R., 2010,
  Phys.\ Rev.\ D, 82, 043518


\bibitem[Kofman, Gnedin \& Bahcall 1993]{Kofman:1993ag}
  Kofman L.A., Gnedin N.Y., Bahcall N.A., 1993,
  ApJ.,  { 413}, 1

\bibitem[Lee 2014]{Lee:2014maa}
  Lee S., 2014,
  Eur.\ Phys.\ J.\ C, {74}, 11,  3146

\bibitem[Linde~1984]{Linde:1984ir}
  Linde A.D., 1984,
  Rep.\ Prog.\ Phys.,  {47}, 925

\bibitem[Matarrese \& Terranova~1996]{Matarrese:1995sb}
  Matarrese S., Terranova D., 1996,
  MNRAS, {283}, 400

\bibitem[Matsubara 1995]{Matsubara95}
Matsubara T., 1995, Prog.\ Theor.\ Phys., 94, 1151

\bibitem[Matsubara 2008a]{Matsubara:2007wj}
  Matsubara T., 2008a,
  Phys.\ Rev.\ D, 77, 063530

\bibitem[Matsubara 2008b]{Matsubara:2008wx}
  Matsubara T., 2008b,
  Phys.\ Rev.\ D,  78, 083519

\bibitem[Mo \& White 1996]{Mo:1995cs}
  Mo H.J., White S.D.M., 1996,
  MNRAS, 282, 347

\bibitem[Moutarde {\it et al.}~(1991)]{Moutarde91}
Moutarde F., Alimi J.M., Bouchet F.R., Pellat R., Raman A., 1991, ApJ, 382, 377

\bibitem[Mukhanov 2005]{Mukhanov}
Mukhanov V., 2005, Physical foundations of cosmology, Cambridge University Press,
Cambridge,  UK, 421 pp


\bibitem[Netterfield C.\ B.\ et al.\ (Boomerang Collaboration)]{Netterfield:2001yq}
Netterfield C.\ B.\ et al.\ (Boomerang Collaboration), 2002, ApJ, 571, 604


\bibitem[Novikov 1970]{Novikov}
Novikov E. A., 1970, Sov.\ Phys.\ J.\ Exp.\ Theor.\ Phys., 30, 512


\bibitem[Peebles 1980]{Peebleslambda}
Peebles P.J.E., 1980, The Large-Scale Structure of the Universe,
Princeton Univ.\ Press, Princeton, NJ

\bibitem[Peebles 1989]{Peebles89}
Peebles P.J.E., 1989,
 ApJ, 344, L53

\bibitem[Planck Collaboration XVI 2014]{Ade:2013zuv}
  Planck Collaboration XVI, 2014,
  A\&A,  {571}, A16

\bibitem[Podvigina {\it et al.}~2015]{Podvigina2015}
Podvigina O., Zheligovsky V., Frisch U., 2015,
J.\ Comput.\ Phys., submitted



\bibitem[Rampf 2012]{Rampfrecursion12}
Rampf C., 2012, J.\ Cosmol.\ Astropart.\ Phys., 1212, 004

\bibitem[Rampf 2013]{Rampf:2013dxa}
  Rampf C., 2013,
  Phys.\ Rev.\ D, 89, 063509

\bibitem[Rampf \& Buchert (2012)]{RampfBuchert}
Rampf C., Buchert T., 2012, J.\ Cosmol.\ Astropart.\ Phys., 1206, 021

\bibitem[Rampf \& Rigopoulos (2013)]{Rampfzelrel}
Rampf C., Rigopoulos G., 2013, MNRAS, 430, L54

\bibitem[Rampf \& Wiegand (2014)]{Rampfwiegand}
Rampf C., Wiegand A., 2014,
  Phys.\ Rev.\ D, {90}, 12, 123503

\bibitem[Rampf \& Wong 2012]{Rampf:2012xb}
  Rampf C., Wong Y.Y.Y., 2012,
  J. Cosmol. Astropart. Phys., 1206, 018

\bibitem[Sahni \& Shandarin (1996)]{Sahni}
Sahni S., Shandarin V., 1996,
 MNRAS, 282, 641 

\bibitem[Vlah, Seljak \& Baldauf 2014]{Vlah:2014nta}
  Vlah Z., Seljak U., Baldauf T., 2014,
  Phys.\ Rev.\ D, {91}, 2,  023508

\bibitem[Zel'dovich (1970)]{Zel'dovich}
Zel'dovich Ya.B., 1970, A\&A, 5, 84

\bibitem[Zheligovsky \& Frisch 2014]{ZheligFrisch2014}
Zheligovsky V., Frisch U., 2014,  J. Fluid Mech., 749, 404

\bibitem[Zheligovsky, Podvigina \& Frisch 2010]{ZheligFrisch2010}
Zheligovsky V., Podvigina O., Frisch U., 2010, J.~Comput.~Phys., 229, 5043

\end{thebibliography}
\end{document}